\documentclass[aps,prd,nofootinbib,12pt,reprint,longbibliography,superscriptaddress,floatfix,floats,tightenlines]{revtex4-2}

\usepackage{graphicx}
\usepackage{epstopdf}
\usepackage{amsmath}
\usepackage{amssymb}
\usepackage{amsfonts}
\usepackage{amsthm}
\usepackage{color}
\usepackage{array}
\usepackage{verbatim}
\usepackage{url}
\usepackage{graphicx}
\usepackage{braket}
\usepackage{epsfig}

\usepackage[
      colorlinks=true,
      linkcolor=blue,
      urlcolor=magenta,
      filecolor=blue,
      citecolor=red,
      pdfstartview=FitV,
      pdftitle={},
      pdfauthor={},
      pdfsubject={},
      pdfkeywords={},
      pdfpagemode={},
      bookmarksopen=true
	  ]{hyperref}
\usepackage{orcidlink}
\usepackage{subfigure}
\usepackage[normalem]{ulem} 
\usepackage{cancel} 
 
\usepackage{soul} 

\newcommand{\abs}[1]{\left|#1\right|}

\newcommand{\figref}[1]{Fig.~\ref{#1}}

\def\frac#1#2{{#1\over #2}}

\def\d{{\mathrm{d}}}
\def\be{\begin{equation}}
\def\ee{\end{equation}}
\def\ba{\begin{eqnarray}}
\def\ea{\end{eqnarray}}

\definecolor{armygreen}{rgb}{0.29, 0.33, 0.13}
\definecolor{darkspringgreen}{rgb}{0.09, 0.45, 0.27}

\numberwithin{equation}{section}

\begin{document}

\title{Gravitational atoms in the braneworld scenario}

\author{Sunil Singh Bohra\orcidlink{0009-0007-8257-9245}}
\email{sunilsinghbohra87@gmail.com}
\affiliation{Center For Theoretical Physics, Jamia Millia Islamia, New Delhi---110025, India}

\author{Subhodeep Sarkar\orcidlink{0000-0001-9015-8837}}
\email[Corresponding author: ]{subhodeep.sarkar1@gmail.com}
\affiliation{Center For Theoretical Physics, Jamia Millia Islamia, New Delhi---110025, India}
\affiliation{Indian Institute of Information Technology, Allahabad, Prayagraj, Uttar Pradesh---211015, India}

\author{Anjan Ananda Sen\orcidlink{0000-0001-9615-4909}}
\email{aasen@jmi.ac.in}
\affiliation{Center For Theoretical Physics, Jamia Millia Islamia, New Delhi---110025, India}

\begin{abstract} 
{In the Randall-Sundrum (RS) II braneworld scenario, general relativity (GR) is modified by adding an extra dimension such that it is indistinguishable from GR in the weak gravity limit. However, such modifications may leave a mark in the strong field regime. We therefore analyze massive scalar perturbations around rotating black holes in the RS II model. Unlike black holes in GR, these braneworld black holes carry a tidal charge that contains information about the extra spatial dimension, and the rotation parameter for such black holes can exceed unity. Through the method of continued fractions, we investigate the quasinormal mode spectra, and the superradiant instabilities associated with the existence of quasibound states, that is, gravitational atoms. In comparison to the four-dimensional Kerr black hole, we report distinctive signatures of the tidal charge and the rotation parameter, which manifest as signals of the extra dimension, on both the fundamental quasinormal mode and the formation of gravitational atoms. These findings offer insights into testing modifications to GR and detecting ultralight bosonic particles around black holes.}

\end{abstract}

\maketitle

\section{Introduction}\label{introduction}
The first direct detection of gravitational waves emitted during the merger of binary black holes and neutron stars \cite{LIGOScientific:2016aoc,LIGOScientific:2017vwq,LIGOScientific:2018mvr}, along with the recent measurements related to the shadows of supermassive black hole candidates likes M87$^*$ and Sgr A$^*$ located at the center of the Messier 87 and the Milky Way galaxies respectively \cite{EventHorizonTelescope:2019dse,EventHorizonTelescope:2019ths,EventHorizonTelescope:2021bee,EventHorizonTelescope:2021srq,EventHorizonTelescope:2023fox,EventHorizonTelescope:2022wkp,EventHorizonTelescope:2022wok} as well as the observation of relativistic effects in the orbits of stars around Sgr A$^*$ \cite{GRAVITY:2018ofz,Do:2019txf,GRAVITY:2020gka}, has led to a renaissance in testing gravity in the strong field regime \cite{Dreyer:2003bv,Berti:2018cxi,Berti:2018vdi,Yagi:2016jml,Hioki:2009na,Bambi:2013nla,Bambi:2015kza,Cunha:2018acu,Narayan:2019imo,Gralla:2019xty,EventHorizonTelescope:2020qrl,Bambi:2019tjh,GRAVITY:2019zin}. To understand classical gravitational interactions, one usually turns to Einstein's general theory of relativity (GR) which is extremely successful as far as solar system experiments, or weak gravitational fields, are concerned \cite{Will:2005yc,Will:2010uh,Will:2014kxa,Will:2018bme}. These recent breakthroughs have now confirmed that the predictions of GR agree with observations to an unprecedented degree even in the strong field regime \cite{GRAVITY:2018ofz,Do:2019txf,GRAVITY:2020gka,LIGOScientific:2016vlm,LIGOScientific:2016lio,LIGOScientific:2017bnn,LIGOScientific:2018dkp,LIGOScientific:2019fpa,LIGOScientific:2020tif,LIGOScientific:2021sio,EventHorizonTelescope:2019ggy,EventHorizonTelescope:2021dqv,EventHorizonTelescope:2022xqj}, adding enormous heft to decades of progress made in this direction \cite{Berti:2004bd,Cardoso:2014uka,Berti:2015itd,Barack:2018yly}. In the near future, pulsar timing arrays and space-borne gravitational wave detectors are expected to put GR to even more stringent tests \cite{Yunes:2013dva,Cannizzaro:2023mgc,Amaro-Seoane:2012vvq,Berti:2004bd,Berti:2019xgr,Barausse:2020rsu,LISA:2022kgy,TianQin:2015yph}. But of course, this whole enterprise is not without its caveats and one must exercise due caution while interpreting the results \cite{Abramowicz:2002vt,Cardoso:2016rao,Berti:2015itd,Cardoso:2019rvt,Mizuno:2018lxz,Gralla:2020pra}.

Moreover, in spite of its tremendous success across various orders of the length scale, there are both theoretical and observational aspects of general relativity and black holes (BHs) that motivate us to consider alternatives to Einstein's theory. Such modified theories of gravity are usually invoked to address issues like the existence of spacetime singularities \cite{Penrose:1964wq,Hawking:1976ra,Christodoulou:1991yfa}, the existence of Cauchy horizons and the breakdown of determinism in GR \cite{Ori:1991zz,Dafermos:2003wr,Bhattacharjee:2016zof,Cardoso:2017soq,Dias:2019ery,Rahman:2019uwf,Rahman:2020guv,Rahman:2020evx,Bhattacharjee:2020gbo,Bhattacharjee:2020nul}, the information loss paradox \cite{Hawking:1975vcx,Chakraborty:2017pmn,Mathur:2009hf}, explain the accelerated expansion of the universe \cite{SupernovaCosmologyProject:1998vns,SupernovaSearchTeam:1998fmf,Sahni:1999gb,Padmanabhan:2002ji,Peebles:2002gy}, or the behavior of galactic rotation curves \cite{Rubin:1978kmz,Bertone:2016nfn}. But when it comes to modifying gravity, we seem to be limited only by our imagination \cite{Nojiri:2006ri,Clifton:2011jh,Berti:2015itd,Nojiri:2017ncd,Shankaranarayanan:2022wbx,Odintsov:2023weg}. However, such alternative theories must be at par with GR when it comes to satisfying tests of gravity, both in the local and strong field regime, which is a tall order. We now know that it is indeed possible to consistently modify the Einstein-Hilbert action such that we end up with theories which can address one or more of these issues while keeping the local physics unchanged. Some popular alternatives to GR include $f(R)$ theories \cite{Nojiri:2010wj,DeFelice:2010aj}, Lanczos-Lovelock models \cite{Padmanabhan:2013xyr}, bimetric gravity \cite{Hassan:2011zd,Babichev:2013pfa,Rahman:2023swz}, Horndeski \cite{Horndeski:1974wa,Sotiriou:2013qea,Babichev:2016rlq} and generalized Proca theories \cite{Heisenberg:2014rta,DeFelice:2016cri,Rahman:2018fgy}.

In the present work, we focus on one such modification to Einstein's theory which incorporates the presence of a warped extra spatial dimension\footnote{The existence of different kinds of extra spatial dimensions have been invoked by physicists in variety of contexts over the years, starting from the works of Kaluza and Klein \cite{Overduin:1997sri} in their attempt to unify classical electromagnetism and general relativity. We refer the reader to \cite{Raychaudhuri:2016kth} for a complete account.}, the so-called braneworld scenario \cite{Randall:1999ee,Randall:1999vf,Chamblin:1999by,Chamblin:2000ra,Shiromizu:1999wj,Sasaki:1999mi,Harko:2004ui,Aliev:2004ds,Garriga:1999yh,Giddings:2000mu,Maartens:2000fg,Maartens:2001jx,Dadhich:2000am,Aliev:2005bi,Aliev:2006qp,Aliev:2009cg,Aliev:2013jqz,Bhattacharya:2001wq,Bento:2002kp,Bento:2002np,Mukhopadhyaya:2002jn,Kim:2003pc,Kanti:2004nr,Perez-Lorenzana:2005fzz,Maartens:2010ar,Raychaudhuri:2016kth}, and try to infer its imprints in the strong field regime by studying the behavior of massive scalar field perturbations around a rotating black hole solution of the said modified theory.

Here we specifically consider the Randall Sundrum braneworld scenario, where we model the universe as a four dimensional ($4D$) spacetime (or a $3$-\-brane) embedded in a five dimensional ($5D$) anti--de Sitter spacetime called the bulk, the extra dimension being spacelike and warped \cite{Randall:1999ee,Randall:1999vf}. In such models, the matter fields are confined to the brane and only gravitational interactions can probe the bulk. The braneworld paradigm was initially proposed as a solution to the hierarchy problem in particle physics \cite{Randall:1999ee}. However, Randall and Sundrum proposed a second model, colloquially known as the RS-\-II model \cite{Randall:1999vf}, where the observable universe is a $3$-\-brane with a positive tension embedded in the $5D$ bulk spacetime with a warped noncompact extra dimension and a negative cosmological constant. It is particularly interesting to note that such a model can give rise to the familiar Newtonian gravitational potential on the $3$-\-brane \cite{Randall:1999vf}. It was also demonstrated that the such a theory is able to reproduce the features of $4D$ Einstein gravity in the low energy limit \cite{Shiromizu:1999wj,Sasaki:1999mi,Harko:2004ui,Aliev:2004ds,Garriga:1999yh,Giddings:2000mu} and numerical black hole solutions in the bulk were explored in \cite{Chamblin:2000ra} (also see \cite{Nakas:2020sey, Nakas:2021srr, Nakas:2023yhj}). {Interestingly enough, the RS-\-II model can also give rise to analytical black hole solutions on the brane \cite{Dadhich:2000am,Aliev:2005bi} that at first sight look superficially familiar to the well known black hole solutions in general relativity.}

{In fact, the rotating braneworld BH metric resembles that of the Kerr-Newman (KN) black hole \cite{Aliev:2005bi}. But the solution is far from trivial since, unlike the KN solution, the braneworld solution corresponds to a vacuum solution. Moreover, the {tidal charge}, bearing the signature of the extra dimension, can take on both positive and negative values (unlike the electric charge).} It is also important to stress that unlike KN BHs of general relativity, the braneworld BH can be \emph{superspinning}, that is, they can possess angular momentum greater than the BH mass. Note that the black holes in our universe are supposed to be electrically neutral. But even then electrically charged black holes are routinely studied because they serve as theoretical laboratories to probe various classical and quantum aspects of gravity \cite{Adamo:2014baa}. Therefore, by using the braneworld BH solution, one can often directly leverage existing techniques to investigate the consequences of the braneworld scenario in the context of gravitational interactions. Such an endeavor is important because it will complement the popular program to search for extra dimensions through particle physics experiments which may be inadequate to probe the presence of the noncompact extra dimension built into the RS-II model \cite{Chung:2000rg}. Against this backdrop, workers have studied the imprint of the tidal charge on lensing and BH shadows \cite{Bin-Nun:2009hct,Bin-Nun:2010exl,Amarilla:2011fx,Banerjee:2019nnj,Banerjee:2019sae,Vagnozzi:2019apd,Neves:2020doc,Banerjee:2022jog}, BH perturbations and gravitational waves \cite{Shen:2006pa,Toshmatov:2016bsb,Chowdhury:2018pre,Mishra:2021waw,Mishra:2023kng,deOliveira:2020lzp,Biswas:2021gvq,Chakraborty:2017qve,Visinelli:2017bny,Rahman:2018oso,Dey:2020lhq,Dey:2020pth,Chakraborty:2022zlq,Chakravarti:2018vlt,Rahman:2022fay,Mukherjee:2022pwd,Ghosh:2021mrv,Mishra:2020jlw}, and have explored various other aspects \cite{Kotrlova:2008xs,Abdujabbarov:2009az,Zakharov:2018awx,Zhang:2019hob,Banerjee:2021aln,deOliveira:2017gxs,deOliveira:2018kcq,deOliveira:2019tlk,deOliveira:2020jha,Biswas:2022wah} as well. 

In this work, we choose to explore two hitherto unexplored avenues: (i) the quasinormal mode spectra of massive scalar perturbations, and (ii) the existence of quasibound states and the associated superradiant instability, or the gravitational atom, in the braneworld scenario.  We focus on the regime $\mu M < 1$ such that one obtains boson condensates around the black hole, forming a scalar gravitational atom, $\mu$ being the mass of the scalar field and $M$ being the BH mass that sets a characteristic length scale of the problem.

{To the best of our knowledge, the existence of gravitational atoms in the context of braneworld black holes has not been examined before, and this work represents the first step in a program to study the implications of the presence of a noncompact extra dimension on the behavior of boson clouds in binary black hole systems.}

 The paper is organized as follows. In Sec. \ref{braneBH} we introduce the rotating braneworld BH metric and discuss the region of the parameter space under exploration. In Sec. \ref{wave_en}, we discuss the wave equation governing massive scalar perturbation and elucidate the boundary conditions required for studying quasinormal modes (QNMs) and quasibound states (QBSs). We then discuss the numerical method in Sec.\ref{leaver} and delineate a strategy to solve the continued fraction using a simple root finding algorithm that guarantees results up to the desired degree of accuracy. We present our results in Sec. \ref{results} where we first analyze the stability of the braneworld BH under massive scalar perturbations, highlighting interesting aspects of the QNM spectrum, followed by a thorough analysis of the quasibound states and the associated superradiant instability. Finally, we conclude with a few remarks in Sec. \ref{remarks}.

\emph{Notations and conventions:} We set the fundamental constants $G$ and $c$ to unity. Throughout this paper, we will use the mostly positive signature convention, such that the Minkowski spacetime will have the metric $\eta_{\mu \nu}=\mathrm{diag}(-1,1,1,1)$. In our numerical computations, we set the characteristic length scale given by the BH mass $M$ to unity.
\section{Rotating black hole in the braneworld scenario} \label{braneBH}
{We had remarked that the rotating braneworld BH resembles the Kerr-Newman BH, and to understand this aspect we must look at how one arrives at the braneworld black hole metric which is a solution to the (effective) Einstein field equations on the brane \cite{Shiromizu:1999wj,Sasaki:1999mi, Harko:2004ui,Aliev:2004ds}. So, in order to construct black hole solutions localized on the brane, one starts with the assumption that the $5D$ Einstein field equations are satisfied by the bulk spacetime. Then, by using an appropriate projector to the brane, and the Gauss-Codazzi relations, one can figure out the $4D$ Einstein field equations on the brane.  In fact, if the bulk spacetime is empty and there are no matter fields present on the brane, then the effective gravitational field equations on the brane are given by \cite{Shiromizu:1999wj,Sasaki:1999mi,Harko:2004ui,Aliev:2004ds}
\begin{equation}
	^{(4)}R_{\mu \nu} = -E_{\mu \nu}, \label{effective_field_equations} \quad ^{(4)}R^{\mu}_{\mu} = E^{\mu}_{\mu} = 0,
\end{equation}
where $^{(4)}R_{\mu \nu}$ is the $4D$ Ricci tensor and $E_{\mu \nu}$ is traceless electric part of the $5D$ Weyl tensor. Therefore, from \eqref{effective_field_equations} it is clear that it is the bulk Weyl tensor that ushers in the modifications to the vacuum Einstein field equations due to the presence of the extra spatial dimension \cite{Shiromizu:1999wj,Aliev:2004ds,Maartens:2001jx}. Now, one can show that $E_{\mu \nu}$ is divergenceless as well if one considers a vacuum brane \cite{Aliev:2004ds}. Therefore, the effective vacuum braneworld field equations closely resemble those of the Einstein-Maxwell system. Dadhich \textit{et al.} \cite{Dadhich:2000am} used this observation to consistently map the Reissner-Nordst\"{o}m solution in GR to an exact static spherically symmetric black hole solution localized on a brane, and soon afterward the technique was generalized to construct a stationary and axisymmetric solution of the vacuum braneworld field equations describing a charged rotating black hole localized on a $3$-brane \cite{Aliev:2005bi}, the charge being an induced \emph{tidal charge}, inherited from the $5D$ Weyl tensor, encoding nonlocal gravitational effects from the higher dimensional bulk spacetime.}

Formally, the line element for a rotating black hole in the second Randall-Sundrum (RS-II) braneworld scenario \cite{Aliev:2005bi,Aliev:2006qp,Aliev:2009cg}, with mass $M$ and angular momentum $J \equiv a M$  in the usual Boyer-
Lindquist coordinates, is given by,
\begin{align}\label{metric}
	ds^2 = & - \dfrac{\Delta}{\Sigma} \left( dt - a \sin^2 \theta {d \phi}\right)^2 + \Sigma \left( \dfrac{dr^2}{\Delta} + \d \theta^2 \right) \nonumber \\
	       & + \dfrac{\sin^2 \theta}{\Sigma}\left( a dt -(r^2 + a^2) d\phi \right)^2,
\end{align}
where the metric functions $\Delta$ and $\Sigma$ have the form,
\begin{align} \label{DeltaSigma}
	\Delta = r^2 + a^2 - 2 M r - \beta, ~ \mathrm{and} ~ \Sigma = r^2 + a^2 \cos^2 \theta,
\end{align}
and $\beta$ is the tidal charge inherited from the bulk Weyl tensor. The tidal charge $\beta$ appears in the metric even though there is no electric charge on the brane and its origin lies in nonlocal Coulomb-type effects present in the bulk space \cite{Dadhich:2000am,Aliev:2005bi}. It is important to note that $\beta$ can take on both positive and negative values, and it is evident that for negative values of $\beta$, the line element \eqref{metric} resembles the standard Kerr-Newman solution of the Einstein-Maxwell system in GR\footnote{The reader should note that our convention is different from \cite{Aliev:2005bi,Aliev:2006qp, Aliev:2009cg}. In their notation, $\Delta = r^2 + a^2 - 2 M r + \beta$. So, in their case, a positive value of $\beta$ corresponds to the KN black hole.}, and for $\beta = 0$ we simply recover the celebrated Kerr solution. Therefore, positive values of $\beta$ differentiates this black hole solution from the standard solutions in general relativity and carries the imprint of the extra dimension. In fact, in the context of braneworld models, a positive value of $\beta$ is physically more favorable \cite{Shiromizu:1999wj,Maartens:2000fg,Chamblin:2000ra}. We should also stress that the tidal charge is a property of the spacetime geometry itself and is different from a black hole hair, which is to say that it is similar to how the cosmological constant is a property of an asymptotically de Sitter black hole spacetime and has the same value for all such black holes, whereas a black hole hair like the mass $M$, or electric charge $Q$ can be different for different black holes in the universe. The rotating braneworld black hole is therefore clearly quite distinct from its GR counterpart despite the superficial similarity. Moreover, the presence of the tidal charge $\beta$ permits us to explore those values of the black hole rotation parameter $a$ that are not allowed in GR. This fact is evident from examining the roots of $\Delta(r) = (r-r_{+})(r-r_{-})=0$, viz.,
\begin{align}\label{eventhorizon}
	r_{\pm} = M \pm k,
\end{align}
where $k \equiv \sqrt{M^2 - a^2 + \beta}$.

\begin{figure}
	\centering
	\includegraphics[width=0.85\linewidth]{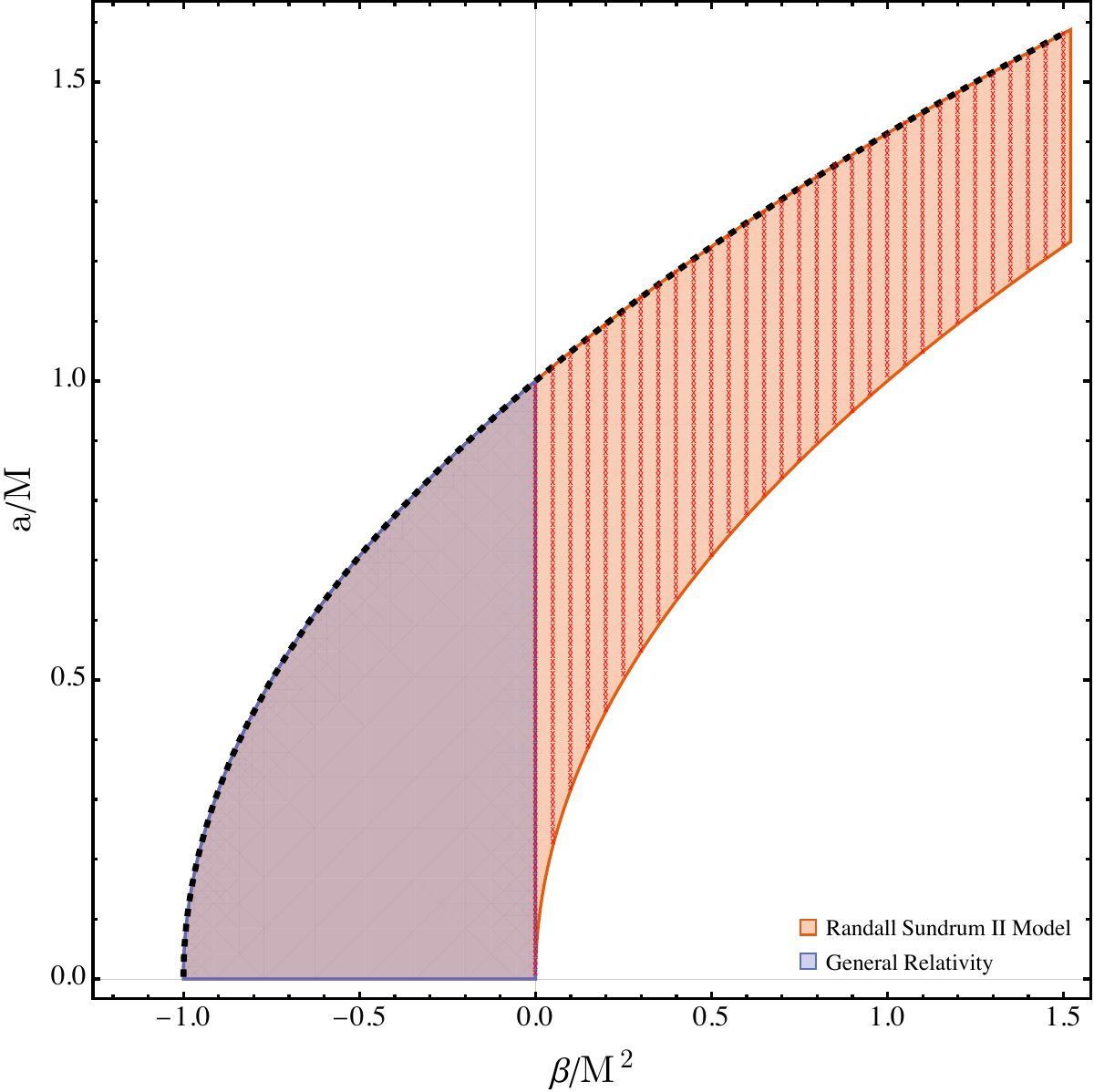}
	\caption{{The space of values of BH spin $a$ and tidal charge $\beta$ which ensure that the rotating braneworld black hole possesses two horizons is shown in orange. The blue region shows the allowed values of $a$ and $\beta$ in the Kerr-Newman spacetime provided $\beta=-Q^2$ is interpreted as the usual electric charge. Although the two solutions are fundamentally different, the braneworld BH being a vacuum solution, they are operationally indistinguishable in the region of overlap ($\beta<0$). The red crosses in the parameter space indicate values of $a$ and $\beta$ sampled to compute the quasinormal modes in the text, and extremal values of $a$ and $\beta$ lie on the thick dotted black curve.}}
	\label{fig:parameter_space}
\end{figure}

The black hole will have an outer event horizon described by the largest root $r_+$ provided,
\begin{align}
	(a/M)^2  \leq 1 + \beta/M^2, \label{lowerlimitbeta}
\end{align}
where the equality in the above relation corresponds to the case where $r_+ = r_- = M$, that is, in the extremal limit when the two horizons will coincide. The violation of \eqref{lowerlimitbeta} results in a spacetime harboring a naked singularity. Now, for the BH to achieve extremality, it is evident that for positive values of $\beta$, we must allow for the possibility $a/M \geq 1$, a situation that is forbidden for black holes in Einstein{\textquotesingle}s general theory of relativity and hence one that is not discussed while studying rotating black holes in GR. Therefore, to explore effects arising out of modifications to general relativity, one has to explore the regime where the rotation parameter $a$ is greater than the mass $M$ of the black hole in presence of a tidal charge $\beta>0$. Furthermore, to guarantee the existence of an inner horizon $r_-$, according to \eqref{eventhorizon}, one must ensure that
\begin{equation}
	M > k \implies \beta < a^2. \label{upperlimitbeta}
\end{equation}
If \eqref{upperlimitbeta} is not satisfied, the spacetime would still describe a rotating black hole but with only one horizon, a geometry whose global structure would be quite different from that of the usual Kerr black hole. We shall therefore restrict our investigation to those black hole geometries which have an inner and outer horizon, that is, focus on values of $\beta$ satisfying,
\begin{equation}
	(a/M)^2-1 \leq \beta/M^2 < (a/M)^2. \label{boundsbeta}
\end{equation}
In \figref{fig:parameter_space}, we have shown the portion of the parameter space spanned by the allowed values $a$ and $\beta$ in the RS-II braneworld scenario (keeping in mind \eqref{boundsbeta}), and compared it with the region described by the KN solution in general relativity. Note that $a$ and $\beta$ are theoretically unbounded from above. We have also indicated the values of $a$ and $\beta$ which satisfy the equality in \eqref{lowerlimitbeta} by a thick dotted black curve. We term this as the \emph{extremal curve}, and later on we shall call the modes corresponding to $a$ and $\beta$ lying close to this curve as near-extremal modes. It is evident that the braneworld black hole provides us with an opportunity to explore a much larger portion of the parameter space compared to GR and hence allows us to investigate the imprint of an otherwise hidden extra dimension on various physical phenomena in the strong field regime.

\section{Massive Scalar Perturbations around the Rotating Braneworld Background} \label{wave_en}

{The goal of the present work is to study the behavior of (massive) scalar field perturbations propagating in the rotating braneworld BH background. Since the seminal work of Vishveshwara \cite{Vishveshwara:1970zz}, quasinormal modes (QNMs), or the characteristic frequencies, of black holes have become ubiquitous to BH physics \cite{Berti:2009kk,Konoplya:2011qq}. These modes are triggered by the presence of perturbing fields near the vicinity of the black hole, or due to perturbations to the background metric itself. They carry unique fingerprints of black hole parameters and spacetime geometry, remaining independent of the initial perturbation. While gravitational perturbations are pivotal from an observational standpoint, the study of massive scalar perturbations is important in its own right \cite{Cardoso:2014uka, Brito:2015oca}. Setting aside the fact that studying the quasinormal mode spectrum of fields of different spins is the first step toward assessing the stability of the BH, massive scalar fields can act as a useful proxy for more realistic baryonic fields. Hence, they act as a useful precursor to full-scale numerical relativity simulations.}

{Furthermore, in asymptotically flat spacetimes, the presence of the scalar field mass fundamentally alters the behavior of the scattering potential at infinity since it now asymptotes to a constant value. This feature leads to an interesting behavior in certain QNM frequencies which can become arbitrarily long-lived, giving rise to the phenomena called \emph{quasiresonance}. Moreover, the massive scalar field potential is now able to trap certain modes since due to the presence of a mass barrier, the radial effective potential acts as a reflecting surface. Such modes, called quasibound states (QBSs), decay exponentially near infinity (in contrast to QNM, even though they still leak away through the horizon) and can be extremely long-lived. In fact, these modes can further extract mass and angular momentum from a rotating BH through the familiar superradiance mechanism \cite{Brito:2015oca} and become \emph{superradiantly unstable} due to successive reflections from the potential barrier, leading to the growth of a scalar condensate outside the BH horizon. Such configurations are similar to the so-called \emph{black hole bombs} \cite{Press:1972zz,Cardoso:2004nk}. In the nonrelativistic limit, these black hole systems carrying a \emph{boson cloud} are also called gravitational atoms since the system loosely resembles that of a hydrogen atom \cite{Arvanitaki:2009fg, Arvanitaki:2010sy,Baumann:2018vus,Baumann:2019ztm,Baumann:2019eav,Baumann:2021fkf,GRAVITY:2023cjt}. These gravitational atoms can be an invaluable tool when it comes to probing the existence of ultralight bosons beyond the Standard Model of particle physics \cite{Arvanitaki:2009fg, Arvanitaki:2010sy,Baumann:2018vus,Baumann:2019ztm,Baumann:2019eav,Baumann:2021fkf,GRAVITY:2023cjt}. Recently, a relativistic framework for studying boson clouds has been proposed as well \cite{Cannizzaro:2023jle}.}

{In this section, we shall review the basics of the wave equation governing massive scalar perturbations and the appropriate boundary conditions to study QNMs and QBSs. The derivation of the concerned equations and boundary condition is operationally identical to that of massive scalar perturbations in the KN background. We therefore only highlight the major steps and results.}

\subsection{The wave equation}

 We start by considering a test field $\psi(t,r,\theta,\phi)$ of mass $\mu$ satisfying the Klein-Gordon (KG) equation.
\begin{equation} \label{klein_gordon_eq}
	\dfrac{1}{\sqrt{-g}} {\partial_\mu}\left(g^{\mu \nu} \sqrt{-g} {\partial _{\nu} \psi}\right)-\mu^{2} \psi=0,
\end{equation}
where $g$ is the determinant of the metric given by \eqref{metric}, and we assume that the test field does not backreact on the background \cite{Berti:2009kk,Konoplya:2011qq}.

Since the metric \eqref{metric} possesses Killing vectors $\partial_t$ and $\partial_\phi$, the wave equation \eqref{klein_gordon_eq} can be solved by the method of separation of variables, using the \emph{ansatz},
\begin{equation}
	\psi (t,r,\theta,\phi)=e^{-i\omega t+i m\phi} S_{lm}{(\theta)} R_{lm}{(r)}, \label{ansatz}
\end{equation}
where  we have introduced the frequency $\omega$ and { the azimuthal number $m$ which is an integer since $\exp{(i m \phi)}$ must remain unchanged under the transformation $\phi \to \phi + 2\pi$.}
Using the ansatz \eqref{ansatz}, we can separate out the KG equation \eqref{klein_gordon_eq} into two coupled ordinary differential equations (ODEs) satisfied by the angular and the radial eigenfunctions, $S_{lm}(\theta)$ and $R_{lm}(r)$ respectively. We therefore find that ${S_{lm}}(\theta)$ satisfies,
\begin{align}
	 & \dfrac{1}{\sin \theta} \dfrac{d}{d \theta}\left(\sin \theta \dfrac{d S_{l m}(\theta)}{d \theta}\right) \nonumber                            \\
	 & + \left(a^{2} (\omega^{2}-\mu^2) \cos^{2} \theta-\dfrac{m^{2}}{\sin ^{2} \theta}+A_{l m}\right) S_{l m}(\theta)=0, \label{angular_equation}
\end{align}
where $A_{lm}$ is the separation constant of the problem, and \eqref{angular_equation} is known as the scalar spheroidal harmonic equation \cite{Berti:2005gp}. {In general, the separation constant is arbitrary, but it takes a discrete set of values labeled by $l,m$ when we demand that $S_{lm}$ must be regular at the poles located at $\theta=0$ and $\theta=\pi$. Then, these solutions are called the scalar spheroidal harmonics, and in the limit $a \to 0$, $S_{lm} \to Y_{lm}$ where  $Y_{lm}$ denote the familiar spherical harmonics, and $A_{lm} \to l(l+1)$ with {$l \geq |m|$} being a non-negative integer. Therefore, $l$ and $m$ can also be used to label the modes.} Note that by introducing a new variable, $u=\cos{\theta}$ where the range $u$ is $-1\leq u \leq 1$, \eqref{angular_equation} can be transformed to the following form,
\begin{align}
	 & \left(1-u^2 \right)\dfrac{d^{{2}} S_{lm}(u)}{du^{{2}}} - 2 u \dfrac{d S_{lm}(u)}{d u}  \nonumber \\
	 & + \left( \bar{\Lambda} + \gamma^2 - \gamma^2 u^2 - \dfrac{m^2}{1-u^2}\right)S_{lm}(u) =0, \label{angular_equation_trigfree}
\end{align}
where we have introduced the constants $\gamma$ and $\bar{\Lambda}$ such that $\gamma = i a \sqrt{\omega^2 -\mu^2}$ and   $A_{lm} = \gamma^2 + \bar{\Lambda}$.  We can also show that the radial eigenfunction $R_{lm}{(r)}$ satisfies,
\begin{equation}
	\Delta  \dfrac{d}{d r} \left( \Delta \dfrac{d R_{lm}(r)}{d r} \right) + U_{lm}(r)R_{lm}(r)=0,\label{radial_eqn}
\end{equation}
where,
\begin{equation}
	U_{lm}(r) =  K^2 - \Delta (\lambda_{lm} + r^2 \mu^2),
\end{equation}
with $\lambda_{lm} = A_{lm} + a^{2}\omega^{2} - 2 a m\omega$, $K =\omega(r^{2}+a^{2}) -a m$ and $\Delta = (r-r_+)(r-r_-)$.

We see that although \eqref{klein_gordon_eq} is separable, the two ODEs \eqref{angular_equation} and \eqref{radial_eqn} that we have obtained are coupled and therefore to solve the eigenvalue problem, we need to solve the two equations simultaneously to determine the separation constant $A_{lm}$ and the eigenfrequencies $\omega$. Furthermore, we have to also supplement Eqs. \eqref{angular_equation} and \eqref{radial_eqn} with suitable boundary conditions (BCs) dictated by the physics of the problem which we shall now discuss.

\subsection{Boundary conditions, quasinormal modes and quasibound states}
To determine the appropriate boundary conditions required to study massive scalar perturbations, we introduce the radial function $\bar{R}_{lm}=\sqrt{r^2+a^2}R_{lm}$ and the tortoise coordinate in the usual manner, viz.,
\begin{equation}
	d r_{*} = \dfrac{r^2 + a^2}{\Delta} dr, \label{tortoise}
\end{equation}
and rewrite the radial equation \eqref{radial_eqn} as,
\begin{equation}
	\dfrac{d^2 \bar{R}_{lm}}{d r_{*}^2} + \bar{U}_{lm} \bar{R}_{lm}=0, \label{Schrodinger}
\end{equation}
where,
\begin{equation}
	\bar{U}_{lm} = \left(\omega - \dfrac{a m }{{r}^2 + a^2}\right)^2-\dfrac{\Delta}{(r^2+a^2)^2}\left( \lambda_{lm} + H(r) \right), \label{schrodinger_pot}
\end{equation}
with,
\begin{equation}
	H(r) = r^2 \mu^2 + \dfrac{(r \Delta)'}{r^2+a^2} - \dfrac{3 \Delta r^2}{(r^2+a^2)^2}.
\end{equation}
Here prime denotes a derivative with respect to $r$, and the resultant equation \eqref{Schrodinger} resembles the Schr\"{o}dinger equation.

Now due to the presence of the event horizon at $r=r_+$, the black hole system is inherently dissipative or nonconservative: waves that fall into the black hole carry energy out of the system and by the virtue of the definition of an event horizon in classical gravity, nothing can emerge from the event horizon as well. Therefore, the scalar waves must satisfy perfectly ingoing boundary conditions at the horizon. To figure out the form of the boundary condition near the horizon, we need to solve \eqref{Schrodinger} asymptotically as $r \to r_+$, or $r_* \to - \infty$, such that $\Delta(r_+) \sim 0$, which in turn implies $\bar{U}_{lm} \sim (\omega - \omega_c)^2$ where
\begin{equation}
	\omega_c = \dfrac{a m }{{r}_+^2 + a^2}. \label{omega_c}
\end{equation}
Noting that near $r_+$, we can write
$$r_* \sim \dfrac{r_+^2 + a^2}{r_+ - r_-} \ln(r-r_+),$$
we find that a solution that is purely ingoing at the event horizon in $v=t+r_*$ coordinates is given by,
\begin{equation}
	R_{lm}(r \to r_+) \sim  e^{-i (\omega-\omega_c )r_*} \sim \left(r-r_{+}\right)^{- i \delta}, \label{rp_ingoing_bc}
\end{equation}
where,
$$\delta = \dfrac{(r_{+}^{2}+a^{2})\omega - a m}{r_{+}-r_{-}}.$$

Now, depending on the boundary condition we choose to impose at $r\to \infty$, we can have different solutions to the eigenvalue problem specified by \eqref{Schrodinger}. In addition to the purely ingoing BC at $r_+$, if we demand that the modes are perfectly outgoing at infinity, we would obtain the \emph{quasinormal modes} (QNMs) of the black hole. However, if the boundary condition is such that the modes decay exponentially near infinity, we get the \emph{quasibound states} (QBSs) of the spacetime. Note that as $r \to \infty$, or $r_* \to \infty$, the potential \eqref{schrodinger_pot} reduces to
\begin{equation}
	\bar{U}_{lm} \sim \omega^2 - \mu^2 + \mu^2\dfrac{ (r_+ + r_-)}{r},
\end{equation}
and hence we can find an asymptotic solution to \eqref{Schrodinger} near infinity to obtain \cite{Konoplya:2004wg, Franzin:2021kvj, Siqueira:2022tbc},
\begin{align}
	R_{lm}(r\to \infty) & \sim r^{i(r_++r_-)\mu^2/2\Omega-1}e^{i \Omega r_*}, \nonumber \\ & \sim  r^{i \rho-1}e^{q r}, \label{radial_bc_infty}
\end{align}
where,
\begin{align}
	\rho   & =\left(\dfrac{2 \omega^{2}-\mu ^{2}}{2 \Omega}\right)(r_{+}+r_{-}),    \\
	\Omega & = \pm \sqrt{\omega^{2}-\mu^{2}},                                       \\
	q      & = i\Omega = {\mp}  \sqrt{\mu^{2}-\omega^{2}}. \label{q}
\end{align}
Note that while obtaining \eqref{radial_bc_infty}, we have consistently taken into account the contribution from the subleading terms near infinity, this is important to ensure the accuracy of the continued fraction method discussed in the next section \cite{Konoplya:2004wg}. It is also interesting to note that \eqref{radial_bc_infty} enables us to study QNMs and QBSs in a unified manner, based on our choice of the sign of $q$ in \eqref{q}. {If $\mathrm{Re}(q)>0$, then the solution diverges near $r \to \infty$, and it is consistent with purely outgoing boundary conditions at infinity. Hence, $\mathrm{Re}(q)>0$ is suitable for studying QNMs. On the other hand, the solution exponentially vanishes near infinity for $\mathrm{Re}(q)<0$, and hence such a choice is suitable for studying QBSs \cite{Dolan:2007mj, Huang:2018qdl}}. The eigenfrequencies $\omega$ that we shall compute are in general complex, that is $\omega = \omega_R + i \omega_I$, unlike the normal mode frequencies of conservative system, and is therefore a hallmark of the dissipative nature of the system due to the loss of energy through the horizon (and also through the infinity for QNMs). The real part of the frequency, $\mathrm{Re}(\omega) \equiv \omega_R$ represents the frequency of oscillation and the imaginary part $\mathrm{Im}(\omega) \equiv \omega_I$ represents the rate of growth or decay of the perturbation. While studying QNMs, for stable perturbations that decay with time, $\mathrm{Im}(\omega)<0$ and  $\abs{\mathrm{Im}(\omega)}$ can be identified with the decay rate. Since we are studying massive scalar field perturbations, the quasibound states may be prone to superradiant instabilities. Such modes are characterized by $\mathrm{Im}(\omega)>0$, and it represents the rate of growth of the instability. Furthermore, the oscillation frequency of these modes is supposed to satisfy \cite{Hod:2009cp,Hod:2014baa,Huang:2016qnk},
\begin{equation}
	0<\mathrm{Re}(\omega)<\mu, ~ \mathrm{and} ~ 0<\mathrm{Re}(\omega)<\omega_c \label{super_cond}.
\end{equation}
The first of the above two conditions ensures that the modes are reflected from the potential barrier at infinity while the second one is the usual condition for superradiance. These two condition together ensure the growth of the superradiant instability and formation of the gravitational atom.

{We end this section by noting that the massive scalar field propagating in the Kerr and Kerr Newman background has received a lot of attention over the last $\sim 50$ years since the pioneering studies of Damour, Deruelle and Ruffini \cite{Damour:1976kh}, Zouros and Eardley \cite{Zouros:1979iw}, and Detweiler \cite{Detweiler:1980uk}, and it continues to be an active area of investigation given its rich phenomenology in light of strong field tests of GR; an excellent description of both the history and the physics of the subject can be found in \cite{Brito:2015oca}. In recent times, it is worth mentioning that the massless scalar and gravitational QNMs in the KN background were studied in \cite{Berti:2005eb}. The QNM spectra of massive scalar fields for Kerr and Kerr Newman black holes were studied in \cite{Konoplya:2006br,Konoplya:2013rxa}. These studies used the method of continued fractions developed by Leaver \cite{Leaver:1985ax}. Studies focusing on the instability of the massive scalar field, or the gravitational atom, in the Kerr background was carried out in \cite{Strafuss:2004qc,Dolan:2007mj,Rosa:2009ei,Hod:2009cp,Dolan:2012yt}, the same for the Kerr Newman black hole was explored in \cite{Furuhashi:2004jk,Huang:2016qnk,Huang:2018qdl,Vieira:2021nha}. Similar studies have been carried out for Kerr-like black holes \cite{Siqueira:2022tbc} and galactic black holes as well \cite{Liu:2022ygf}. But before we move on to study the behavior of massive scalar fields in the context of the braneworld scenario, we discuss the numerical method used to obtain our results in the next section.}

\section{A Numerical Recipe for Leaver's Method of Continued Fractions}
\label{leaver}
Armed with \eqref{angular_equation_trigfree}, \eqref{radial_eqn}, and the boundary conditions \eqref{rp_ingoing_bc} and \eqref{radial_bc_infty}, we now attempt to numerically determine the QNMs and QBSs of the braneworld BH. We have chosen the method of continued fractions proposed by Leaver \cite{Leaver:1985ax} to compute QNMs. The method itself was first used by Jaff\'{e} to compute the electronic spectra of hydrogen molecular ion, and over the years it has been widely used to compute the QNMs and QBSs of various black hole geometries. We shall first write down the necessary recurrence relations {which resemble those of the KN black hole}, and then discuss a strategy based on \cite{Siqueira:2022tbc} to scan the parameter space (c.f., \figref{fig:parameter_space}) in a manner that assures convergence.

\subsection{Radial equation}

We begin by considering the following \emph{ansatz} for the radial differential equation that takes into account the BCs  \eqref{rp_ingoing_bc} and \eqref{radial_bc_infty}, viz.,
\begin{equation}
	R_{lm}{(r)}= e^{i \Omega r} \left(\dfrac{r-r_{+}}{r-r_{-}}\right)^{i \delta} (r-r_{-})^{i \rho}P(r),\label{radial_anstaz_form}
\end{equation}
where,
$$
	P({r})=\sum_{n=0}^{\infty} d_{n}\left(\dfrac{r-r_{+}}{r-r_{-}}\right)^n.
$$
We then plug in \eqref{radial_anstaz_form} into \eqref{radial_eqn} to obtain a three term recurrence relation satisfied by the coefficients $a_n$, namely,
\begin{equation}
	\alpha_{0}^{r} a_{1} + \beta_{0}^{r} a_{0} = 0,\label{radial_recurr_a1a0}
\end{equation}
\begin{equation}
	\label{radial_recurr_rel}
	\alpha_{n}^{r} d_{n+1} + \beta_{n}^{r} d_{n}+\gamma_{n}^{r} d_{n-1}=0, \; \;\; \; \; {n\ge 1}
\end{equation}
with,
\begin{subequations}
	\label{radial_alphabetagama_rel}
	\begin{align}
		 & \alpha_{n}^{r} = n^{2} + (c_{0}+1)n + c_{0},           \\
		 & \beta_{n}^{r} = -2 n^{2} + (c_{1}+2)n + c_{3},         \\
		 & \gamma_{n}^{r} = n^{2}+(c_{2}-3)n + c_{4} - c_{2} + 2,
	\end{align}
\end{subequations}
and,
\begin{subequations}
	\label{radial_c0c1c2c3c4_rel}
	\begin{align}
		c_{0} = & 1+\dfrac{2 i}{(r_{+}-r_{-})}\left(am - \omega(a^{2}+r_{+}^{2}) \right),                             \\
		c_{1} = & -2(c_{0}+1)+ 4 i  r_{+}\Omega + \dfrac{i \mu^{2}(r_{+}+r_{-})}{\Omega},                             \\
		c_{2} = & c_{0}+2-i (r_{+}+r_{-})\left(\dfrac{2 \omega^{2}-\mu^{2}}{\Omega}\right),                           \\
		c_{3} = & c_{0}\left[c_{0} + \dfrac{c_{1}}{2} + i \omega(r_{+}+r_{-})\right]- i \omega(r_{+}+r_{-}) \nonumber \\
		        & + a^{2} \omega^{2}+r_{+}^{2}(2\omega^{2} - \mu^{2})- A_{lm}, \label{c3}                             \\
		c_{4} = & c_{0} -i (r_{+}+r_{-})(c_{0}-1)\omega  - \dfrac{(r_{+}+r_{-})^{2}\mu^{4}}{4\Omega^{2}}   \nonumber  \\
		        & -i (r_{+}+r_{-})(c_{0}+1) \dfrac{(2\omega^{2}-\mu^{2})}{2 \Omega}.\label{c4}
	\end{align}
\end{subequations}
We now divide the \eqref{radial_recurr_rel} by $d_{n}$  and then, after a minor rearrangement, obtain
\begin{equation} \label{radia_recur_to_fraction}
	\dfrac{d_{n}}{d_{n-1}} =  - \dfrac{\gamma_{n}^{r}}{\beta_{n}^{r}+\alpha_{n}^{r}\dfrac{d_{n+1}}{d_{n}}}.
\end{equation}
Now \eqref{radia_recur_to_fraction} can be cast in the form of an infinite continued fraction by substituting the expression for ${d_{n+1}}/{d_{n}}$ [obtained by replacing $n$ with $n+1$ in \eqref{radia_recur_to_fraction}] back into \eqref{radia_recur_to_fraction}, and iterating the process till we obtain,
\begin{equation} \label{infinte_conti_frac1}
	\dfrac{d^{r}_{n}}{d^{r}_{n-1}} = - \dfrac{\gamma_{n}^{r}}{\beta_{n}^{r}-\alpha_{n}^{r}\dfrac{\gamma_{n+1}^{r}}{\beta_{n+1}^{r}- \alpha_{n+1}^{r}\dfrac{\gamma_{n+2}^{r}}{\beta_{n+2}^{r}-\cdots}}}.
\end{equation}

Putting $n=1$ in \eqref{infinte_conti_frac1} and equating it with \eqref{radial_recurr_a1a0} we further get,
\begin{equation}\label{inf_cont_frac}
	0 = \beta_{0}^{r}-\dfrac{\alpha_{0}^{r} \gamma_{1}^{r}}{\beta_{1}^{r}-} \dfrac{\alpha_{1}^{r} \gamma_{2}^{r}}{\beta_{2}^{r}-} \cdots  \dfrac{\alpha_{n}^{r} \gamma_{n+1}^{r}}{\beta_{n+1}^{r}-} \cdots .
\end{equation}
The above continued fraction can be inverted $n$ number of times to yield,
\begin{align}
	\label{inv_cont_frac}
	\beta_{n}^{r} - \dfrac{\alpha_{n}^{r} \gamma_{n+1}^{r}}{\beta_{n+1}^{r}-} \dfrac{\alpha_{n+1}^{r} \gamma_{n+2}^{r}}{\beta_{n+2}^{r}-} \cdots =\dfrac{\alpha_{n-1}^{r} \gamma_{n}^{r}}{\beta_{n-1}^{r}-} \cdots \dfrac{\alpha_{0}^{r}\gamma_{1}^{r}}{\beta_{0}^{r}} .
\end{align}
We note that putting $n=0$ in \eqref{inv_cont_frac} gives back  \eqref{inf_cont_frac}, provided that for $n<0$, $\alpha_n=\beta_n=\gamma_n=0$. If we know the value of the separation constant $A_{lm}$ appearing Eq. \eqref{c3}, we can truncate the continued fraction at a suitably large value $N$ and solve the algebraic equation using an appropriate root finding algorithm to determine the QNM or QBS frequency $\omega$ depending on the sign of $\Omega$ (as discussed in the previous section), thereby solving the radial eigenvalue problem. The infinite continued fraction \eqref{inv_cont_frac} in principle has an infinite number of roots, however numerically, the $n{\mathrm{th}}$ inversion gives the $n{\mathrm{th}}$ stable root. So to study the fundamental mode, we have to put $n=0$ in  \eqref{inv_cont_frac}. The inverse continued fraction is actually more useful in calculating the overtones.

Since we truncate the continued fraction at some $N$, the \emph{remainder} of the series $R_{N} = -{d_{N+1}}/{d_{N}}$ can be approximated following Nollert's prescription \cite{Nollert:1993zz}. Since $R_N$ satisfies the recurrence relation,
\begin{align}
	R_{N} = -\dfrac{\gamma_{N+1}^{r}}{\beta_{N+1}^{r}-\alpha_{N+1}^{r} R_{N+1}},
\end{align}
we can then expand $R_N$ as a power series, viz.,
$$R_{N}=\sum_{k=0}^{\infty} C_{k} N^{-k/2}, $$
and then by equating the coefficients of each power of $\sqrt{N}$ to zero, we obtain the first three coefficients as,
$$ \begin{aligned}
		 & C_{0}= -1,                                                                        \\
		 & C_{1}=\pm \sqrt{- 2 i\, \Omega(r_{+}-r_{-})},                                     \\
		 & C_{2} = \dfrac{3}{4} + 2i \Omega r_{+} + \dfrac{i \mu^{2}(r_{+}+r_{-})}{2\Omega}.
	\end{aligned} $$
Note that the sign of $C_{1}$ is chosen in such a way that Re($C_{1}$) should be positive to ensure convergence. Now for a large value of $N$, the contribution from the terms following $N$ become very small. So, including terms beyond the $N^{\text{th}}$ term should not significantly alter the value of the root the continued fraction. However, rather than simply discarding or selecting an arbitrary value for the remaining part, it is preferable to apply Nollert's prescription since Nollert observed that it is crucial when one attempts to calculate modes whose imaginary parts are much larger than their real part, and as a result improves the overall convergence of Leaver's method.

\subsection{Angular equation}
We need to solve the angular equation \eqref{angular_equation_trigfree} to determine the separation constant $A_{lm}$ by imposing the boundary condition that the function $S_{lm}(u)$ is regular at the poles $u=\pm1$. Taking this into consideration, the series solution for the angular wave-function takes the form,
\begin{equation} \label{ang_soln_anstaz}
	S_{lm}=e^{i \Omega u}(1+u)^{m / 2}(1-u)^{m / 2} \sum_{n=0}^{\infty} c_{n}(1+u)^{n},
\end{equation}
Substituting \eqref{ang_soln_anstaz} into \eqref{angular_equation_trigfree} yields the following three term recurrence relation
\begin{equation} \label{ang_recurr_rel}
	\begin{aligned}
		 & \alpha_{n}^{\theta} c_{n+1} + \beta_{n}^{\theta} c_{n} + \gamma_{n}^{\theta}c_{n-1} = 0,
	\end{aligned}
\end{equation}
\text{where}
\begin{subequations}
	\begin{align}
		\alpha_{n}^{\theta}= & 2(n+1)(n+m+1),                                  \\
		\beta_{n}^{\theta}=  & -n^{2}+n(4a\Omega - 2m - 1) + A_{lm}  \nonumber \\
		                     & + a^{2}\Omega^{2}  + (2a\Omega - m)(m+1),       \\
		\gamma_{n}^{\theta}= & -2 a \Omega(n+m).
	\end{align}
\end{subequations}
Although we can solve both the radial and angular equations on an equal footing using Leaver's method, in practice we have found that it often much more computationally inexpensive to use a suitable library function \cite{Liu:2022ygf} to compute $A_{lm}$. 
\subsection{Implementation}
Since we have transformed the problem of finding out the eigenvalues of the ODEs \eqref{angular_equation_trigfree} and \eqref{radial_eqn} into a problem of finding the roots of an infinite continued fraction \eqref{inv_cont_frac}, we can solve \eqref{inv_cont_frac} to determine the eigenfrequency $\omega$ by specifying the set of values $B =\{a, \beta, \mu, l,m\}$ along with the number of terms to include in the continued fraction $N$ and a suitable guess value $\omega_0$ for the root-finding algorithm. To use \eqref{inv_cont_frac}, we also need to specify $n$, the number of inversions of the continued fraction \eqref{inf_cont_frac}, and since we shall be focusing on the fundamental mode, we set $n=0$. We also set $M=1$ as the characteristic length scale which means that all dimensionful quantities are suitably scaled with respect to $M$ {and made dimensionless}.

To ensure convergence of the mode $\omega(B)$ for a set of values specified by $B$ and $\omega_0$, we adopt the following strategy: we first determine $\omega(B)$ for $N=N_1$, and then increment $N$ by $dN$ and recompute the value of $\omega(B)$. We continue the iteration until $N$ reaches a maximum value $N_{\mathrm{max}}$, or the relative difference between the values of ${\omega(B)}$ from two successive iterations falls below a specified tolerance  $ \epsilon = 10^{- p}$. Symbolically, if $\omega(B;N)$ is the value of the root $\omega(B)$ obtained by keeping $N$ terms in the continued fraction, we break the iteration when
\begin{equation}
	\log_{10} \abs{ 1 - \dfrac{\omega({B};{N+dN})}{\omega({B};{N})}} < - p. \label{convergence_criteria}
\end{equation}
In our calculation, we have set $p=7$, ensuring that our results converge up to at least six decimal places, unless stated otherwise. The values of $\{N_1, dN, N_{\mathrm{max}}\}$ are so chosen as to ensure that the computation finishes in a feasible amount of time on a workstation while ensuring convergence, e.g.: for the $l=m=0$ QNMs, we take, $N_1=100, dN = 100, N_{\max}=5000$. We have also noticed that while computing QNMs, the continued fraction converges for relative small values of $N \sim 300-600$ when the black hole is far from extremality, but for near-extreme configurations it requires a high value of $N\sim 1000-3000$ on average to ensure that the modes converge to six decimal places. Since the scattering problem in black hole spacetimes is inherently dissipative, the eigenvalue spectrum is prone to numerical instabilities arising from rounding-off errors due to machine precision arithmetic \cite{Jaramillo:2020tuu}, therefore it is customary and prudent to perform intermediate calculations using extended precision. In our work we have therefore set the internal precision to at least $4\times\mathrm{MachinePrecision}$. The strategy outlined so far is used in the computation of both the QNMs and QBSs.

In order to determine the quasinormal modes, we scan the parameter space \figref{fig:parameter_space} in the following manner: we fix the values of $l,m$ and the mass $\mu$ of the scalar field and choose a value of the tidal charge $\beta$. Since there is theoretically no restriction on the value of the tidal charge, we restrict ourselves to $\beta \leq \beta_{\mathrm{max}}=1.5$. We then increment the value of $a$ from $0$ to the maximum possible value $a_{\mathrm{max}}=\sqrt{1+ \beta_{\mathrm{max}}}$ in steps of $da = 0.01$, and calculate the corresponding fundamental QNM frequency $\omega$ for those values of $a$ given $\beta$ such that the inequality given by \eqref{boundsbeta} is satisfied. For the first allowed value of $a$ given $\beta$, we use a value approximately equal to the of the fundamental massless scalar quasinormal mode of a Schwarzschild BH or a slowly rotating Kerr BH ($a=0.000001$) as the initial guess value $\omega_0$ for the root-finding algorithm\footnote{The initial guess value, or an approximate value of the fundamental mode of the Schwarzschild or the slowly rotating Kerr black hole can itself be estimated by computing the logarithm of the absolute value of right-hand side of \eqref{inf_cont_frac} over a suitable region of the complex plane, and checking for which value of $\omega = x + i y$, we get a minimum, since this point will lie close to the root of the continued fraction that we are after. The process can be repeated for two values of $N$ to ensure that the minima is not a numerical artifact.}. But for subsequent iterations, we use the value of the mode found in the previous iteration as the new value of $\omega_0$. This is advantageous because near extremality, we found that the convergence of the root-finding algorithm is extremely sensitive to the choice of $\omega_0$; for an ``improper" choice of $\omega_0$, the algorithm may also return a value of $\omega$ that corresponds to an overtone instead of the fundamental mode. We repeat the process described so far for other values of $\beta$ in parallel, and we choose values of $\beta$ lying between $0$ and $\beta_{\mathrm{max}}$ in steps of $d\beta=0.05$. The sampled points are shown in \figref{fig:parameter_space} as red crosses.

To compute the quasibound states and the associated superradiant instability, {we follow a slightly modified approach to scan the parameter space \figref{fig:parameter_space} since the root finding algorithm is extremely sensitive to choice of guess values for nonzero values of $\beta$ and $\mu$. We proceed in two major steps. First, we fix the values of $l,m$ and $\mu$ and then, by using a guess value $\omega_0$ [based on \eqref{super_cond}] such that its real and imaginary parts are $\mathrm{Re}(\omega_0)=0.95 \mu$ and $\mathrm{Im}(\omega_0)=10^{-8}$ respectively, we calculate the QBSs for $a \sim 0$ and $\beta=0$. We now keep $a \sim 0$ fixed and increase the value of $\beta$ by $d \beta =0.01$. Then, using the value of the mode we have just computed as the new guess value, we compute the QBS for $a\sim0$ and $\beta=d \beta$. We then move horizontally by incrementing $\beta$ until we reach $\beta_{\mathrm{max}}=1.5$, keeping $a\sim0$ fixed all the while. This step gives us the first rung of guess values that we shall use in the second major step to scan the parameter space by moving vertically upward, that is, by keeping $\beta$ fixed and incrementing $a$ by $da=0.01$ for each $\beta$ in parallel. In subsequent iterations, we shall use the mode computed in the previous step as new guess value. This process is different from the one employed in computing QNMs in two ways: first, we use a ``customized" guess value for each $\beta$ and second, we end up computing the QBSs even for those geometries for which $\beta>a^2$. However, while plotting the results we discard those values of $a$ and $\beta$ that violate the inequality given by \eqref{boundsbeta}.}

{We are also interested in how the modes behave with respect to variations in $\mu$, focusing on a much smaller set of values of $a$ and $\beta$.} So, now we fix the values of $a, l,m$, and choose a set of values of $\beta$ which satisfies \eqref{boundsbeta}. Then for each pair of $a$ and $\beta$, we calculate the quasibound state starting with $\mu=\mu_{\mathrm{min}}$. We then increment $\mu$ in steps of $\d \mu$ up to $\mu_{\mathrm{max}}$. For the first iteration, while calculating the QBS for $\mu_{\mathrm{min}}$, we set the real and imaginary part of the guess value to $\mathrm{Re}(\omega_0)=0.95 \mu_{min}$ and $\mathrm{Im}(\omega_0)=10^{-8}$ respectively, in accordance with \eqref{super_cond}. In subsequent iterations, we use the value of the QBS computed in the previous step as the new guess value. For quasibound states, the first overtone tends to lie very close to the fundamental mode, and hence the root-finding algorithm might be highly sensitive to the choice of the guess value. In our approach, this difficulty may be ameliorated by choosing a smaller value of $d\mu$. Furthermore, in the regime of the superradiant instability, the imaginary part of the QBS has a very small positive value compare to its real part. Therefore, to ensure that the modes do indeed converge, we have to apply the criteria given by \eqref{convergence_criteria} to both real and imaginary parts of $\omega$ separately and simultaneously. We have observed that the real part converges much faster than the imaginary part and hence requires a stricter test.

Lastly, we have validated our approach by confirming that it is able to reproduce existing results related to the quasinormal modes and quasibound states of Kerr and Kerr-Newman black holes \cite{Dolan:2007mj,Huang:2018qdl,Liu:2022ygf}.

\begin{figure*}[!htb]
	\centering
	\begin{minipage}{0.45\textwidth}
		\centering
		\includegraphics[width=\linewidth]{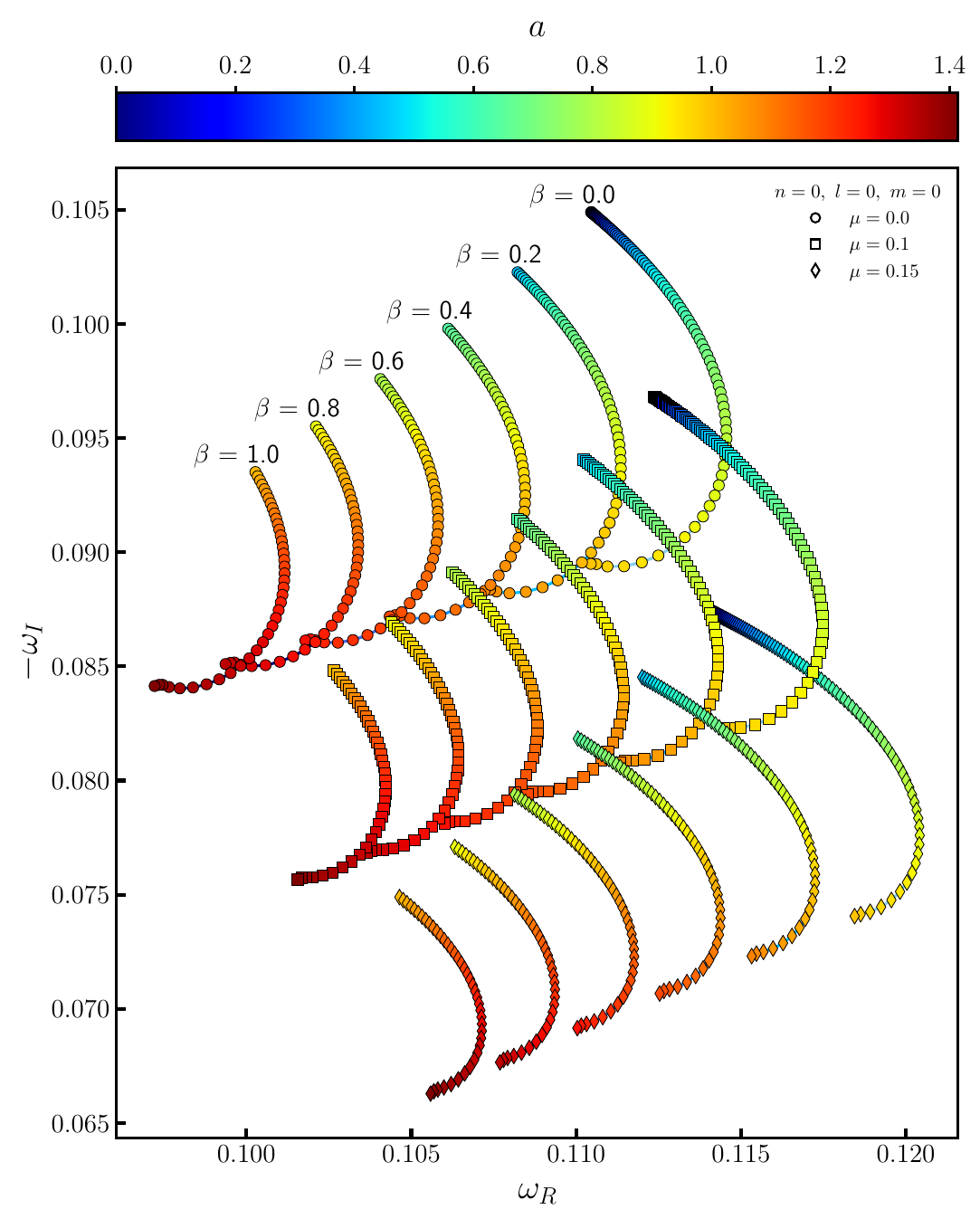}
	\end{minipage}
	\hfill
	\begin{minipage}{0.45\textwidth}
		\centering
		\includegraphics[width=\linewidth]{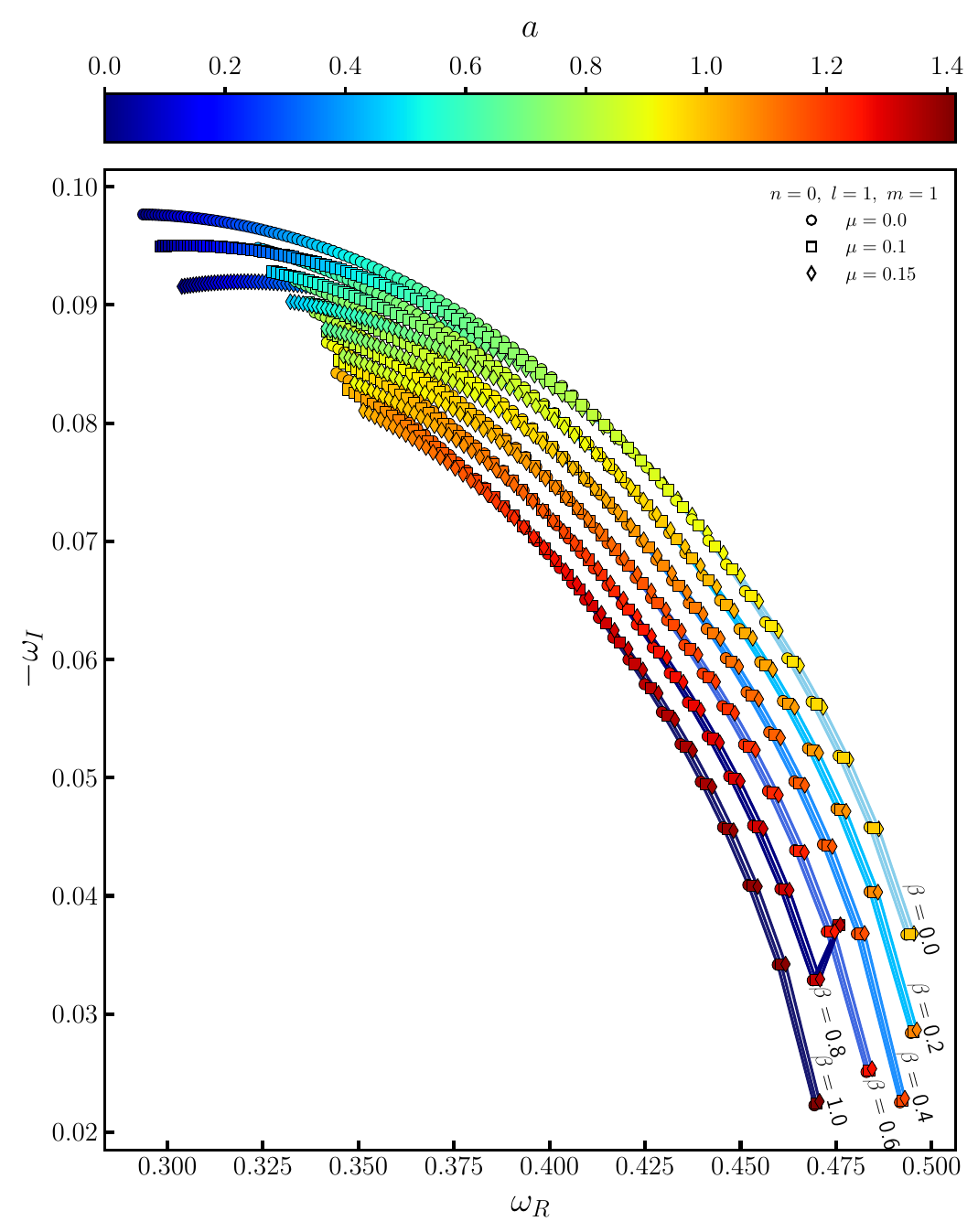}
		\label{fig:qnm_parametric_plotl1m1}
	\end{minipage}
	\vspace{1em} 
	\begin{minipage}{0.45\textwidth}
		\centering
		\includegraphics[width=\linewidth]{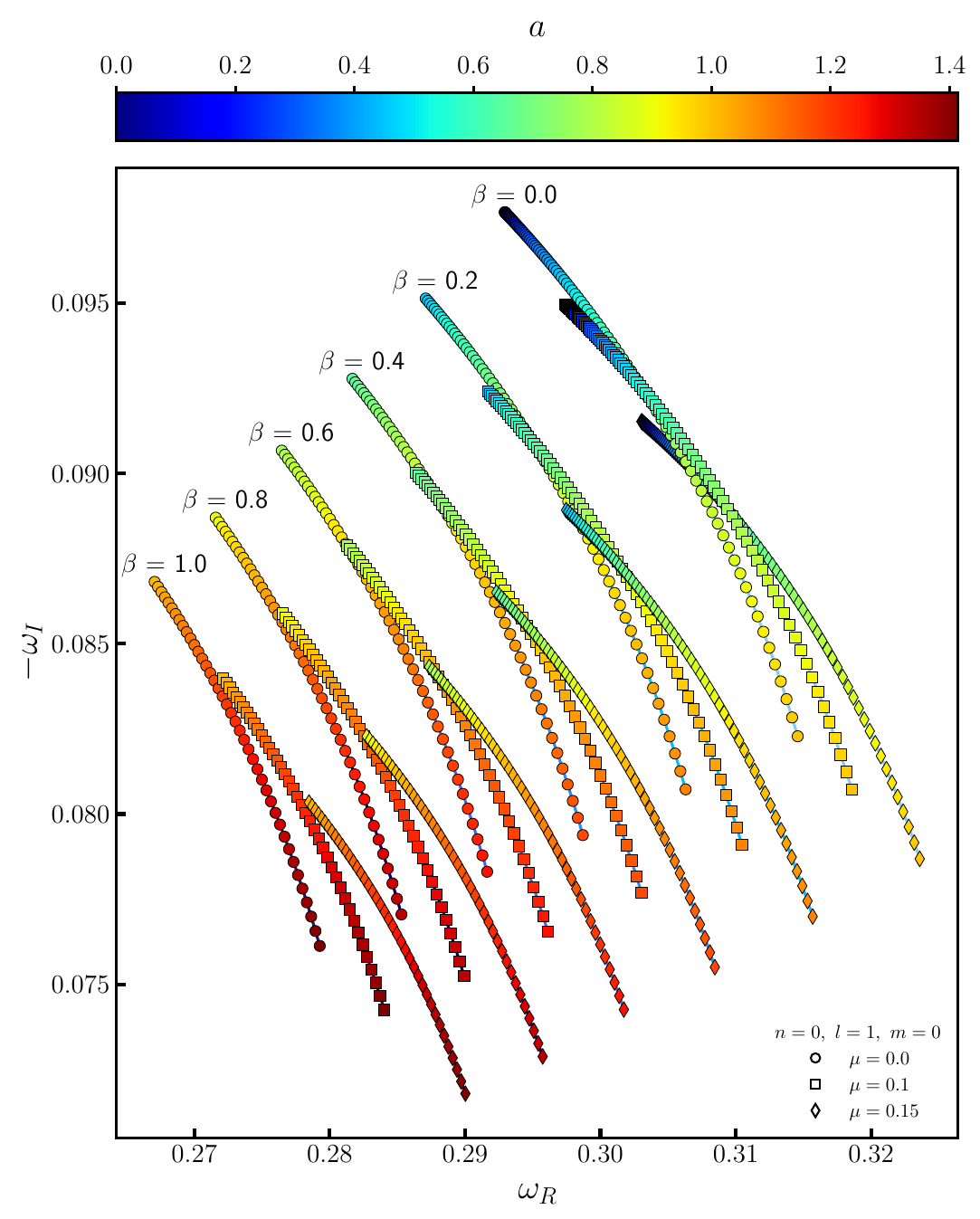}
		\label{fig:qnm_parametric_plotl1m0}
	\end{minipage}
	\hfill
	\begin{minipage}{0.45\textwidth}
		\centering
		\includegraphics[width=\linewidth]{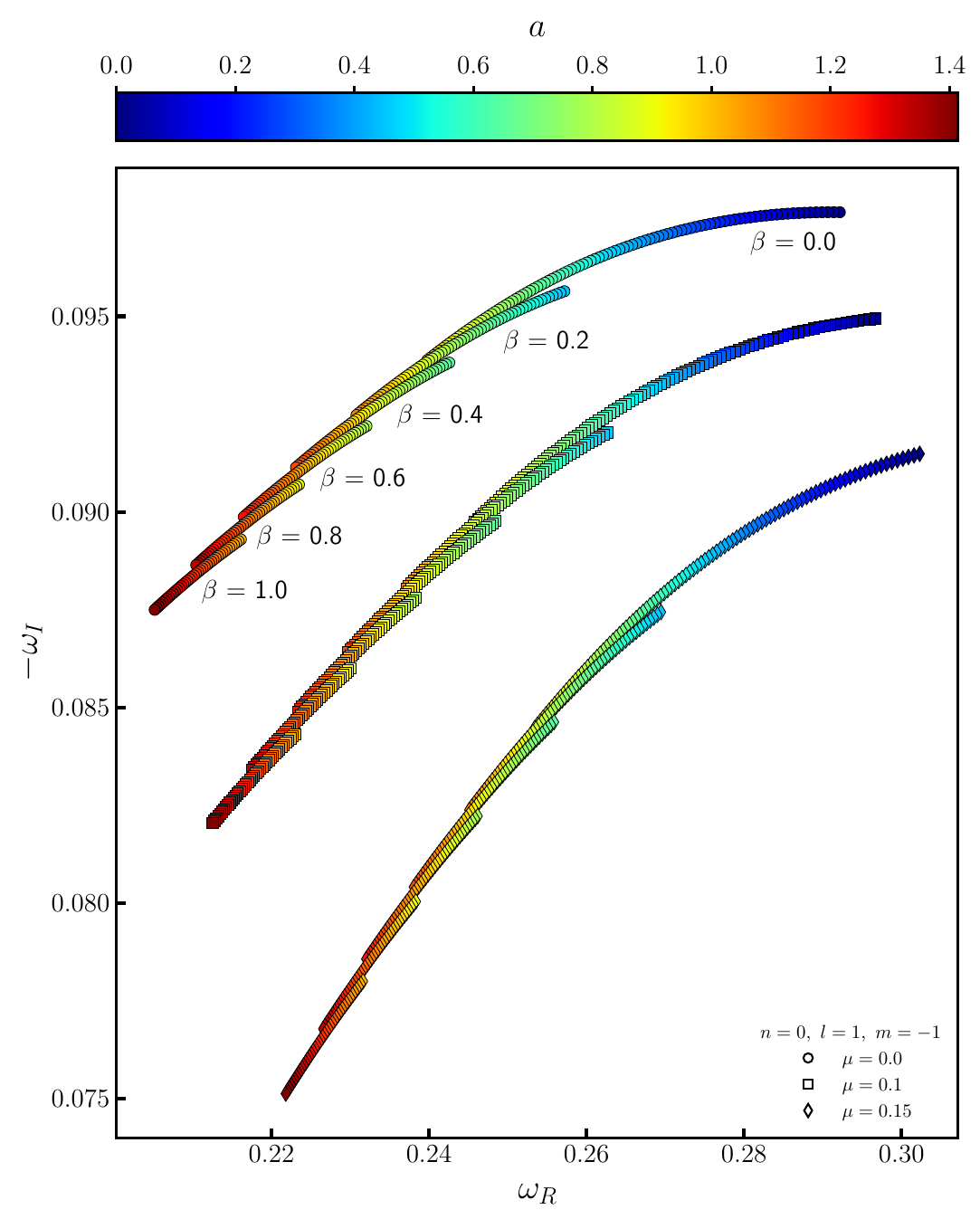}
		\label{fig:qnm_parametric_plotl1m-1}
	\end{minipage}
	\caption{The quasinormal mode spectra of the rotating braneworld black hole for $l=m=0$ (top left), $l=m=1$ (top right), $l=1,m=0$ (bottom left) and $l=1,m=-1$ (bottom right). Each curve in the complex plane is labeled by $\beta$, and the color of each point corresponds to the value of $a$, whereas the value of $\mu$ is indicated by the shape of the marker. In these figures, we have set the characteristic length scale given by the black hole mass $M$ to unity.}
	\label{fig:qnm_parametric_plot}
\end{figure*}

\section{Numerical Results}
\label{results}
In this section, we shall present the results of our numerical explorations, focusing first on the quasinormal modes, and then on the quasibound states and the associated superradiant instability.
\subsection{Quasinormal mode spectra}
\begin{figure*}[!htb]
	\centering
	\minipage{0.33\textwidth}
	\includegraphics[width=\linewidth]{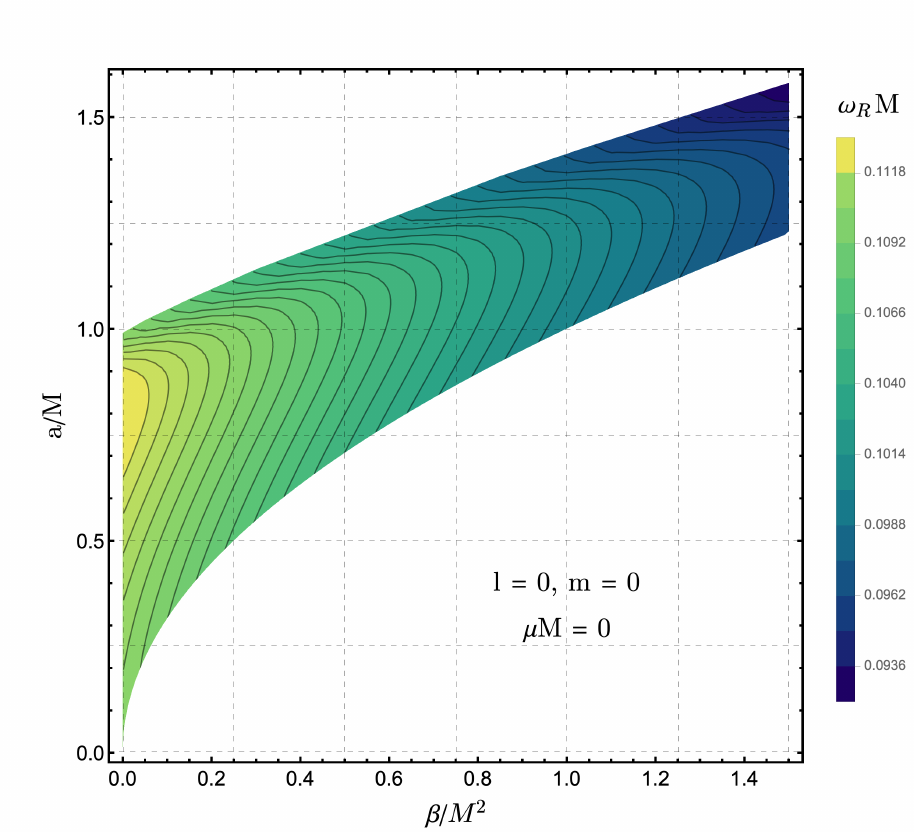}
	\endminipage\hfill
	\minipage{0.33\textwidth}
	\includegraphics[width=\linewidth]{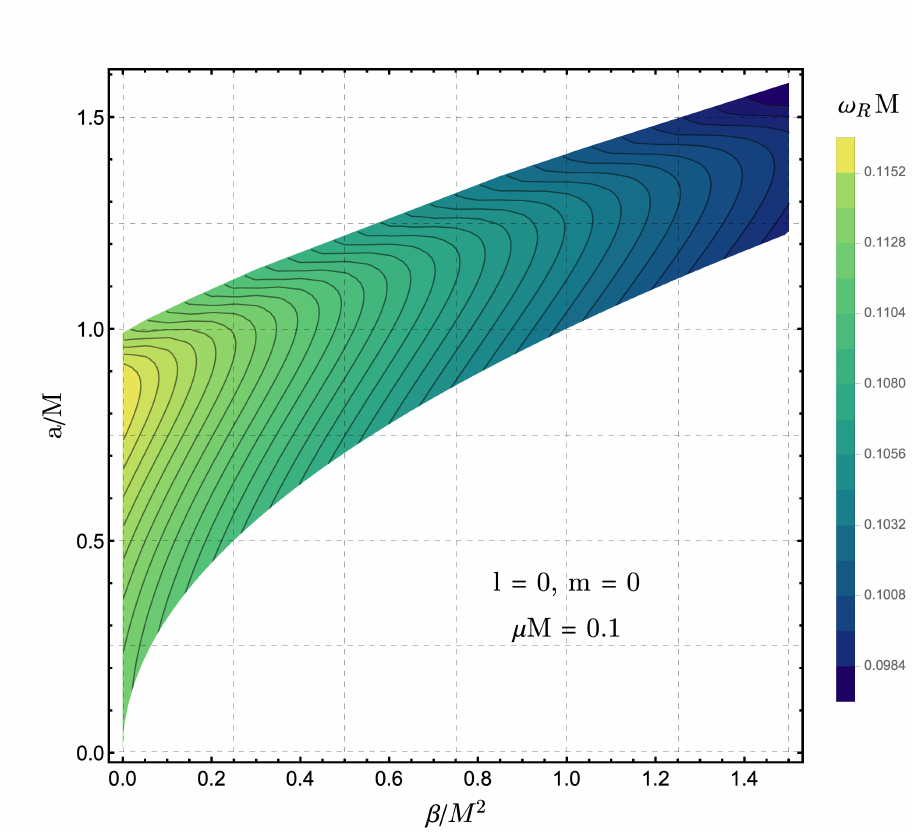}
	\endminipage
	\hfill
	\minipage{0.33\textwidth}
	\includegraphics[width=\linewidth]{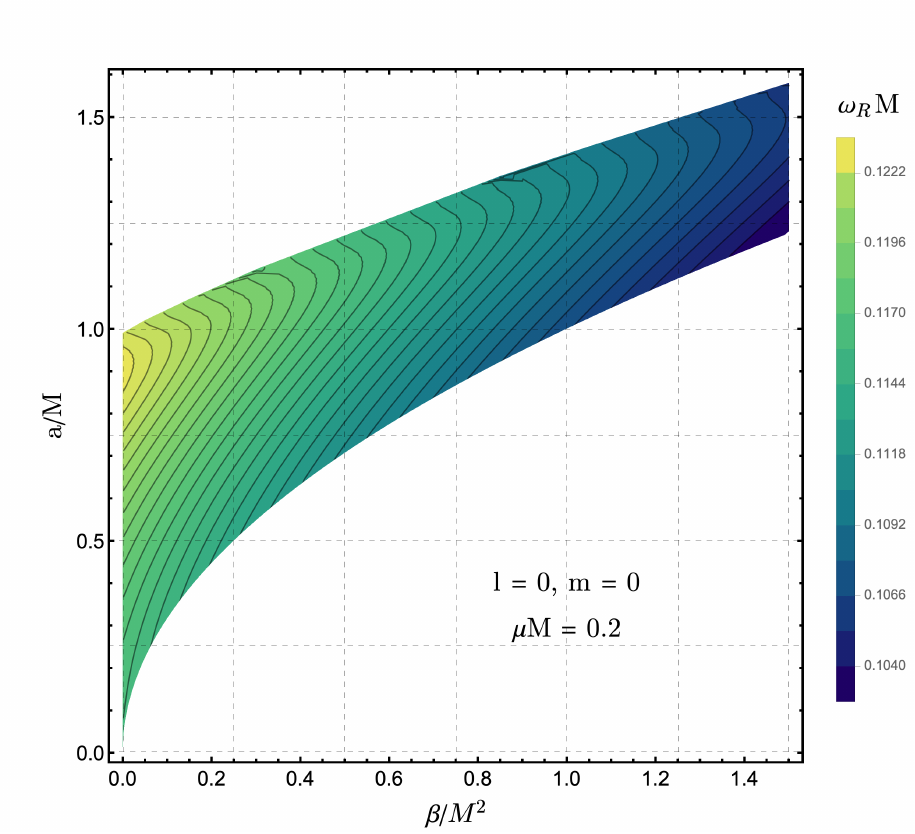}
	\endminipage\hfill
	\minipage{0.33\textwidth}
	\includegraphics[width=\linewidth]{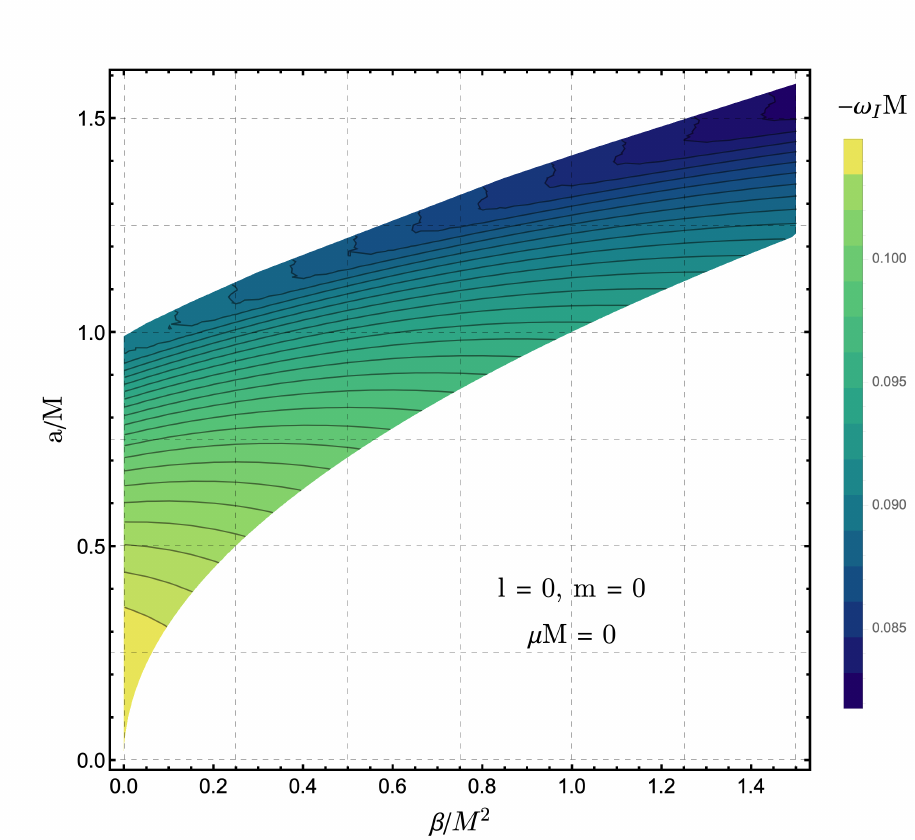}
	\endminipage\hfill
	\minipage{0.33\textwidth}
	\includegraphics[width=\linewidth]{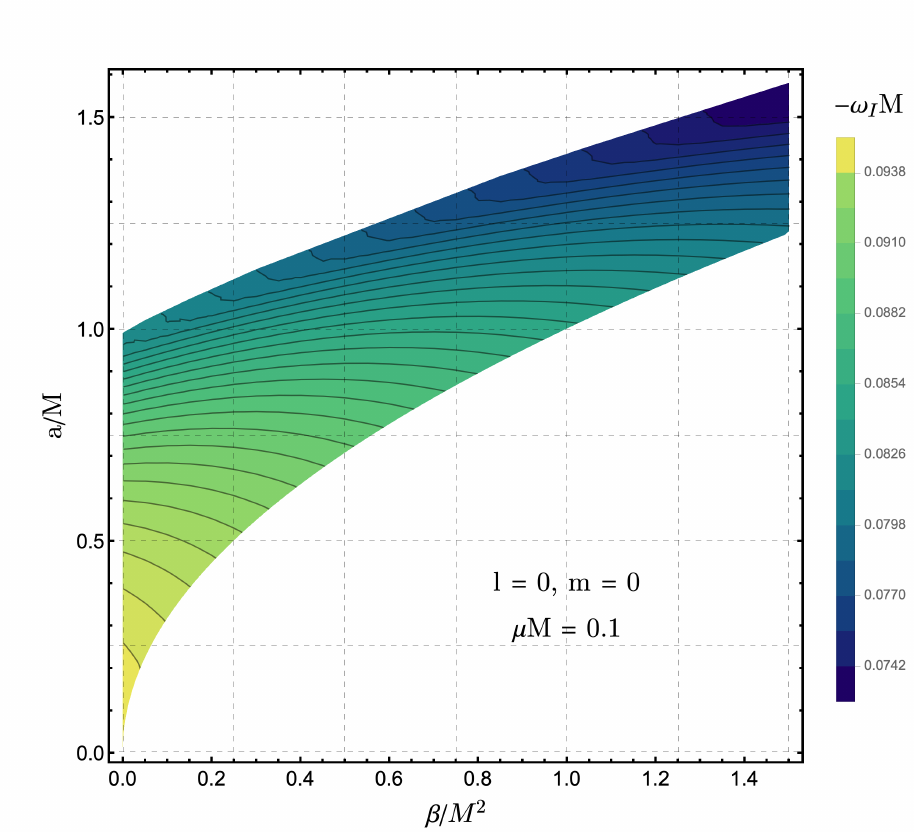}
	\endminipage
	\hfill
	\minipage{0.33\textwidth}
	\includegraphics[width=\linewidth]{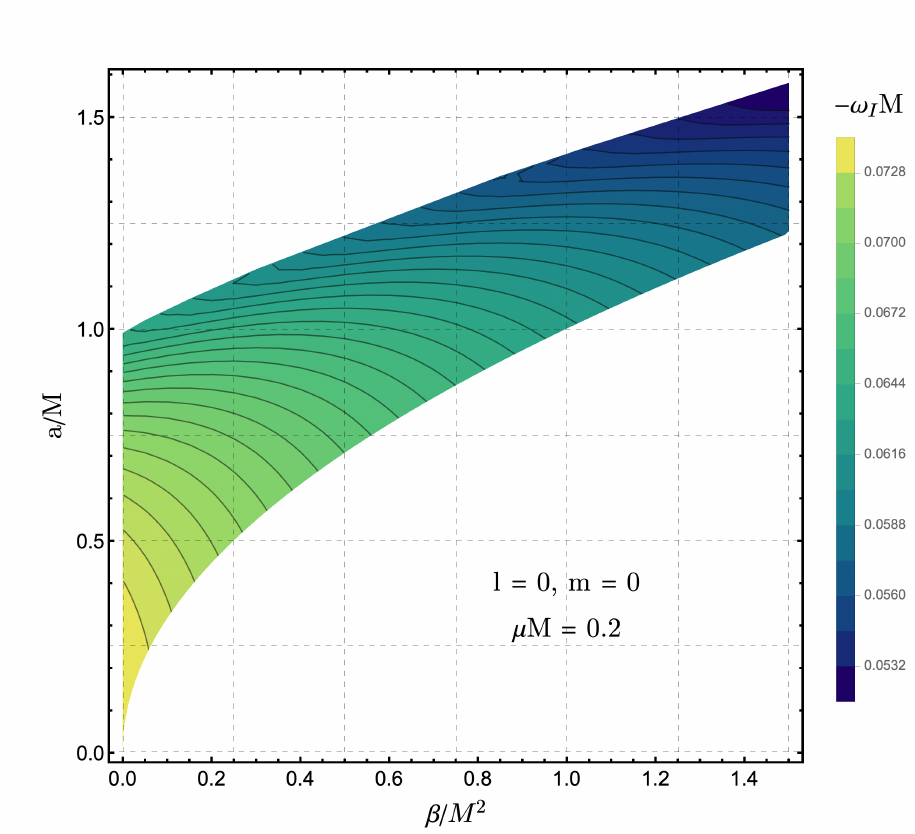}
	\endminipage
	\caption{{The real (top row) and imaginary (bottom row) parts of the scalar quasinormal mode spectra corresponding to the $l=m=0$ mode for $\mu M=0$ (left column), $\mu M=0.1$ (middle column), $\mu M=0.2$ (right column) for the allowed values of $a/M$ and $\beta/M^2$.}}\label{fig:QNM_RealImg_l0}
\end{figure*}

\begin{figure*}[!htb]
	\centering
	\minipage{0.33\textwidth}
	\includegraphics[width=\linewidth]{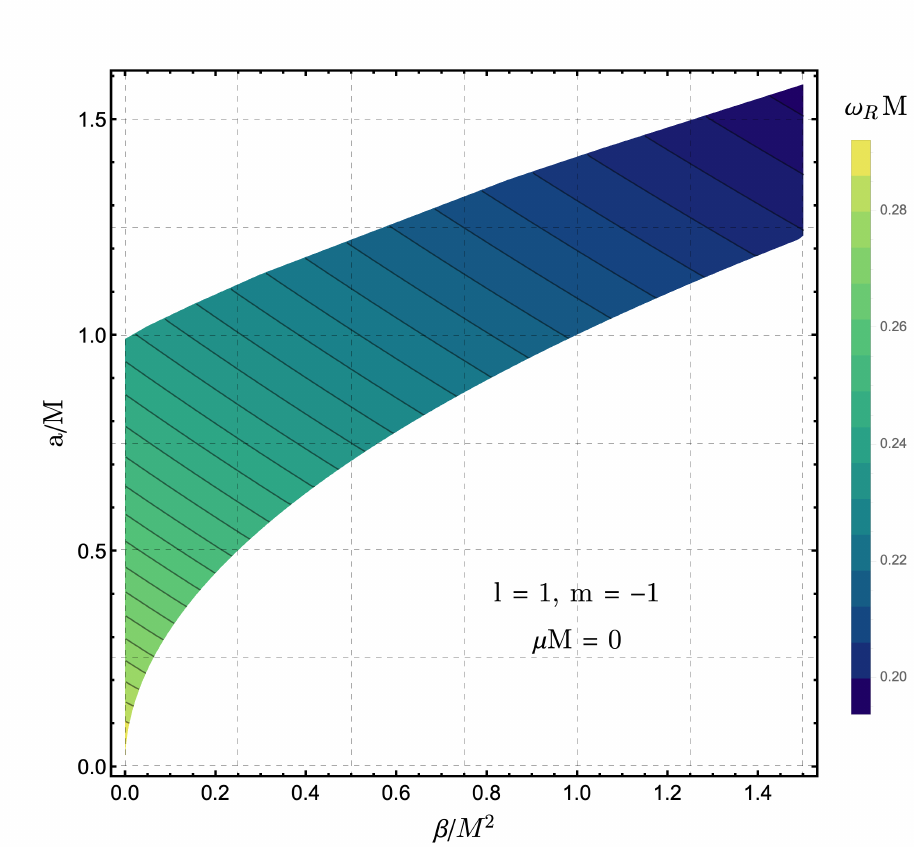}
	\endminipage\hfill
	\minipage{0.33\textwidth}
	\includegraphics[width=\linewidth]{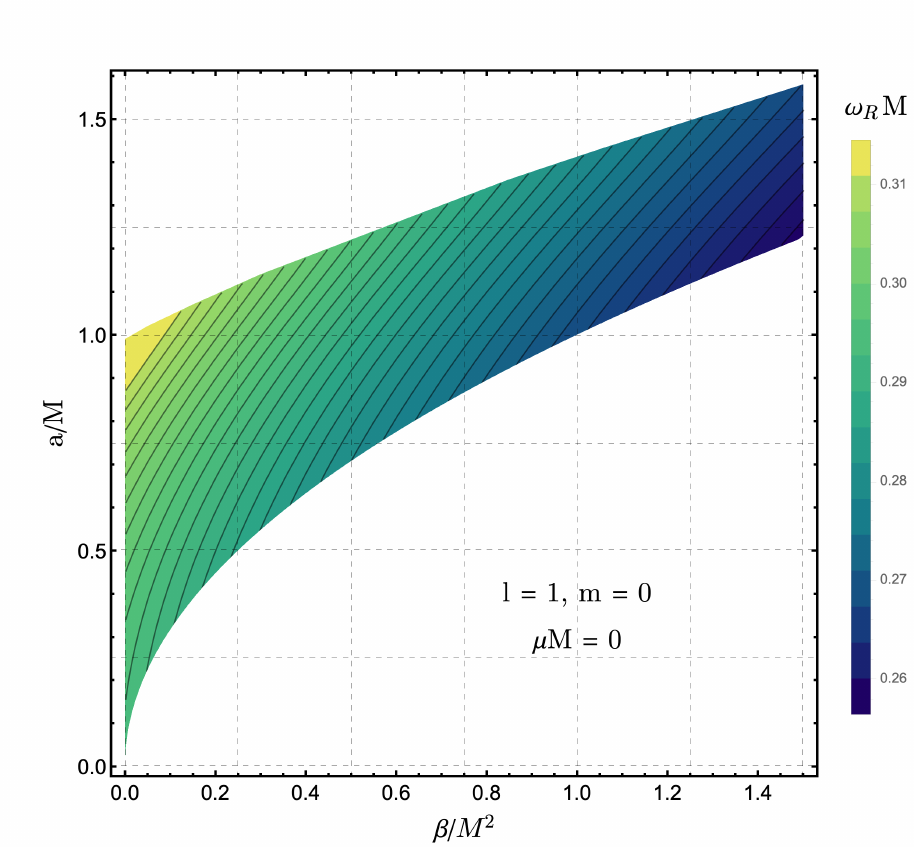}
	\endminipage
	\hfill
	\minipage{0.33\textwidth}
	\includegraphics[width=\linewidth]{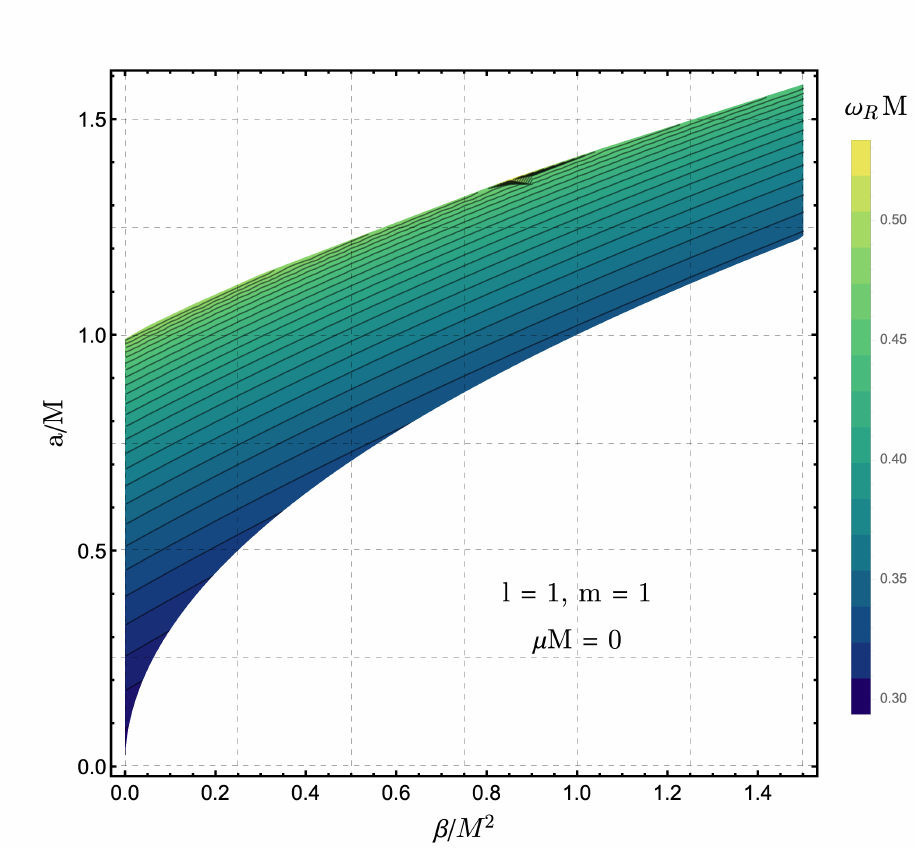}
	\endminipage\hfill
	\minipage{0.33\textwidth}
	\includegraphics[width=\linewidth]{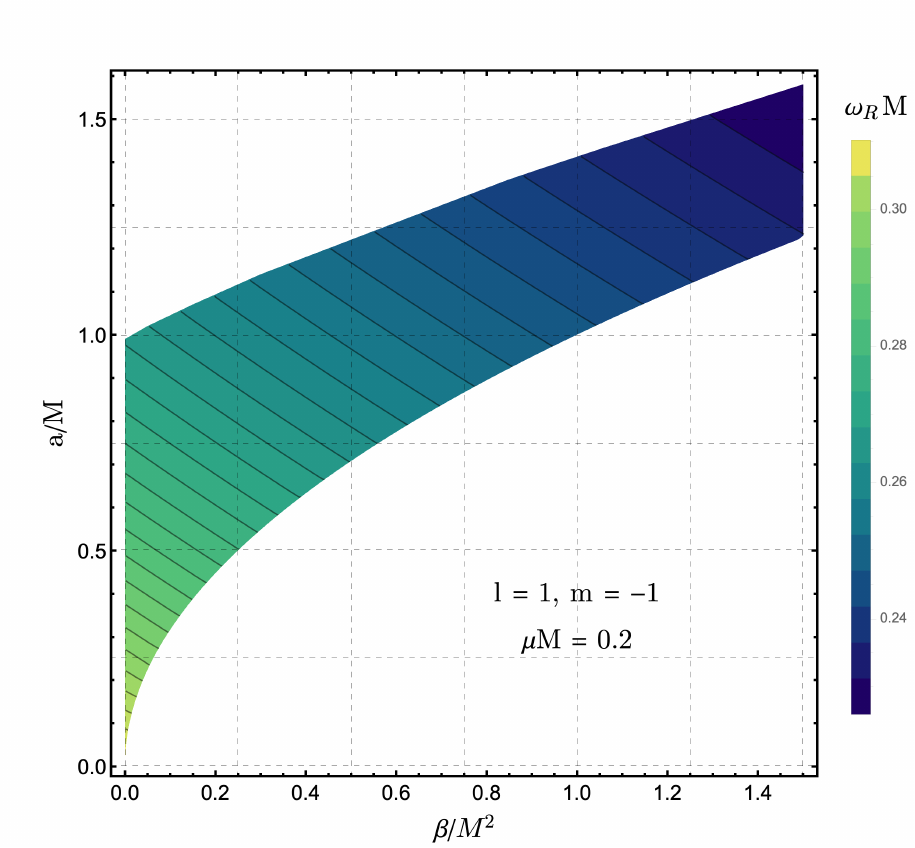}
	\endminipage\hfill
	\minipage{0.33\textwidth}
	\includegraphics[width=\linewidth]{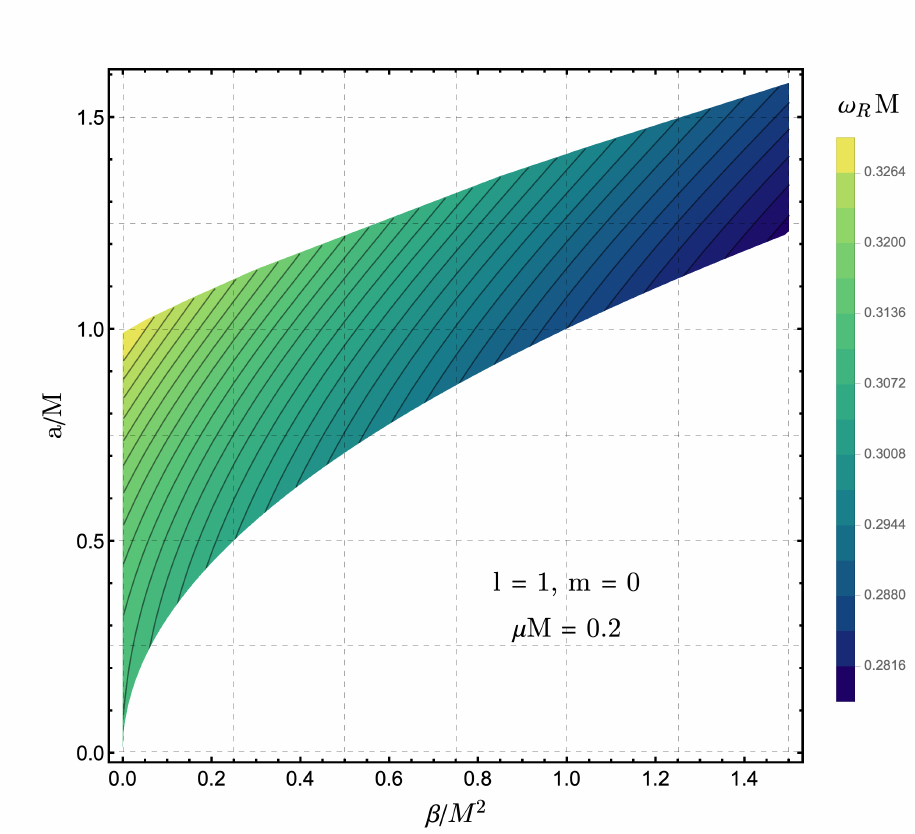}
	\endminipage
	\hfill
	\minipage{0.33\textwidth}
	\includegraphics[width=\linewidth]{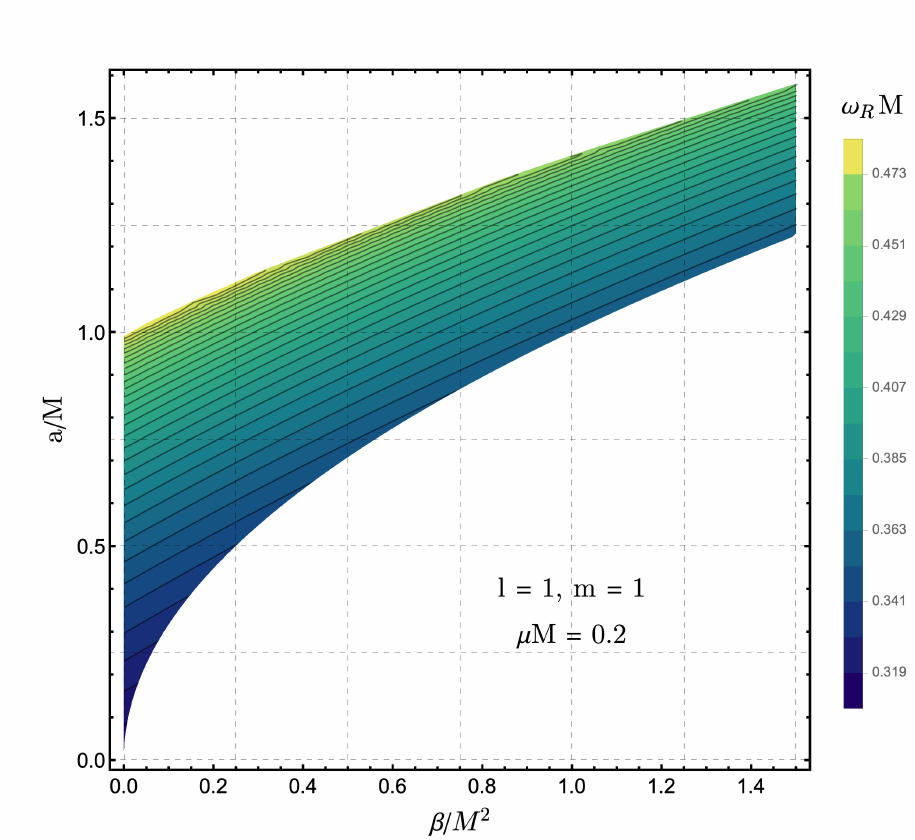}
	\endminipage
	\caption{The real part of the scalar quasinormal mode spectra corresponding to $l=1$ and $m=-1$ (left column), $m=0$ (middle column), $m=1$ (right column) for the allowed values of $a/M$ and $\beta/M^2$ for $\mu M=0$ (top row) and $\mu M=0.2$ (bottom row).}\label{fig:QNM_Real_l1}
\end{figure*}

\begin{figure*}[!htb]
	\centering
	\minipage{0.33\textwidth}
	\includegraphics[width=\linewidth]{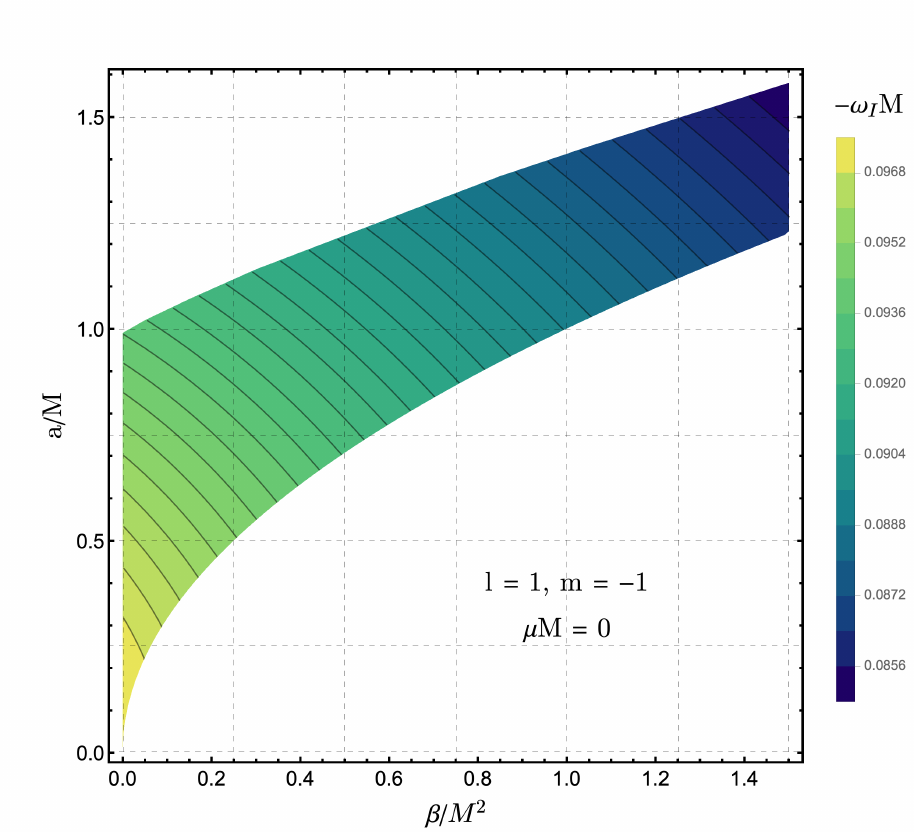}
	\endminipage\hfill
	\minipage{0.33\textwidth}
	\includegraphics[width=\linewidth]{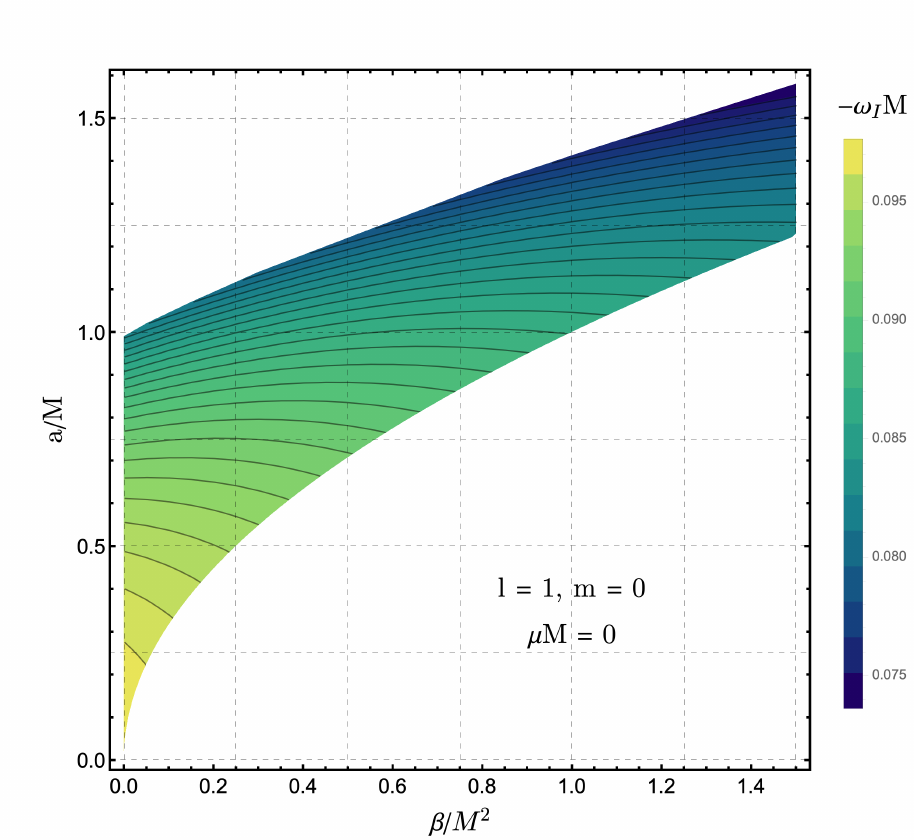}
	\endminipage
	\hfill
	\minipage{0.33\textwidth}
	\includegraphics[width=\linewidth]{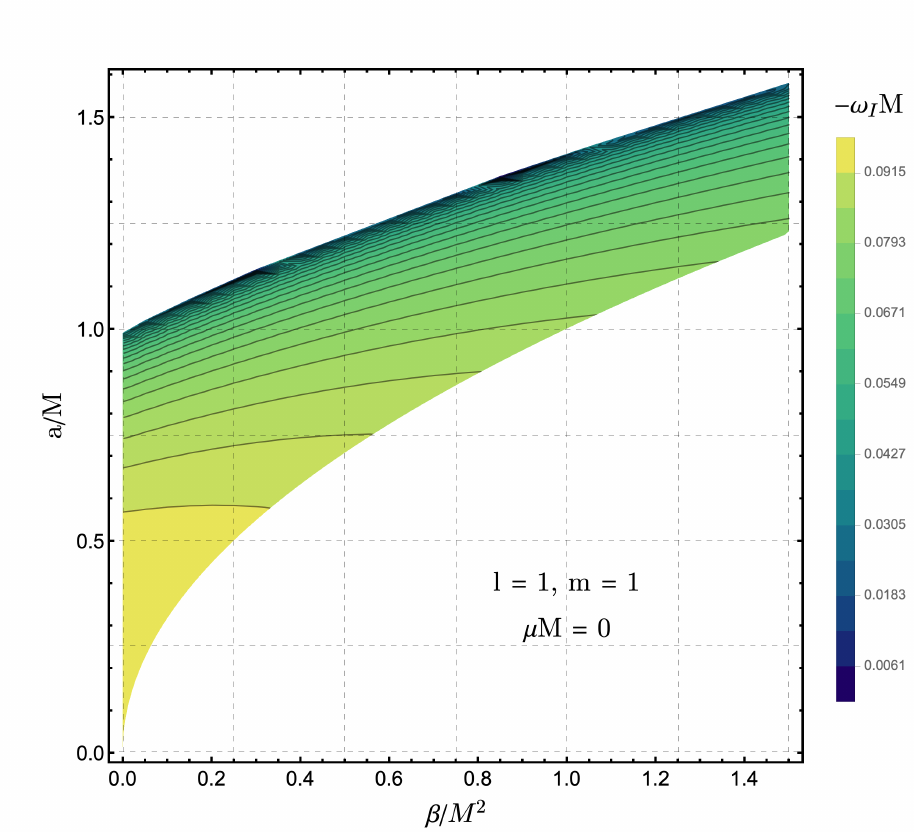}
	\endminipage\hfill
	\minipage{0.33\textwidth}
	\includegraphics[width=\linewidth]{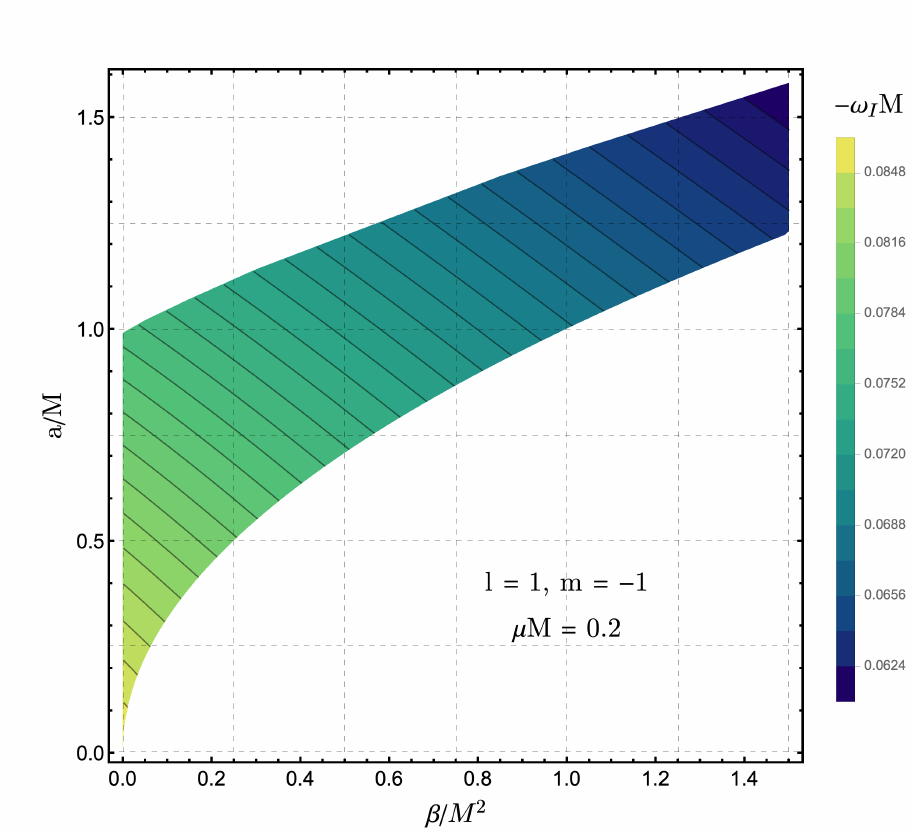}
	\endminipage\hfill
	\minipage{0.33\textwidth}
	\includegraphics[width=\linewidth]{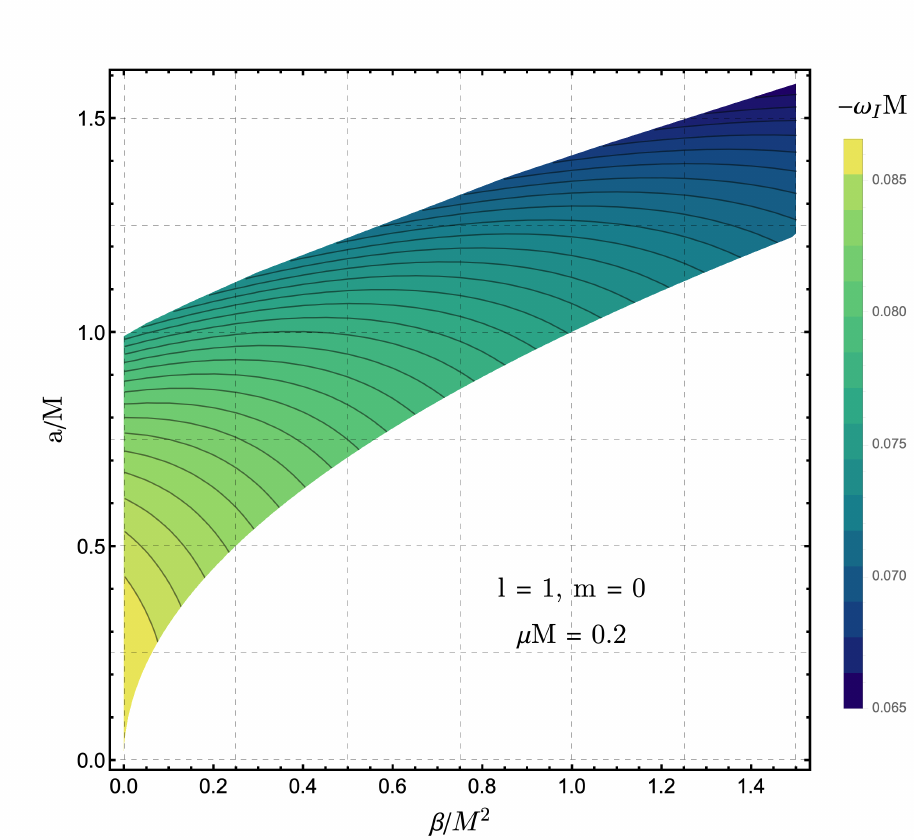}
	\endminipage
	\hfill
	\minipage{0.33\textwidth}
	\includegraphics[width=\linewidth]{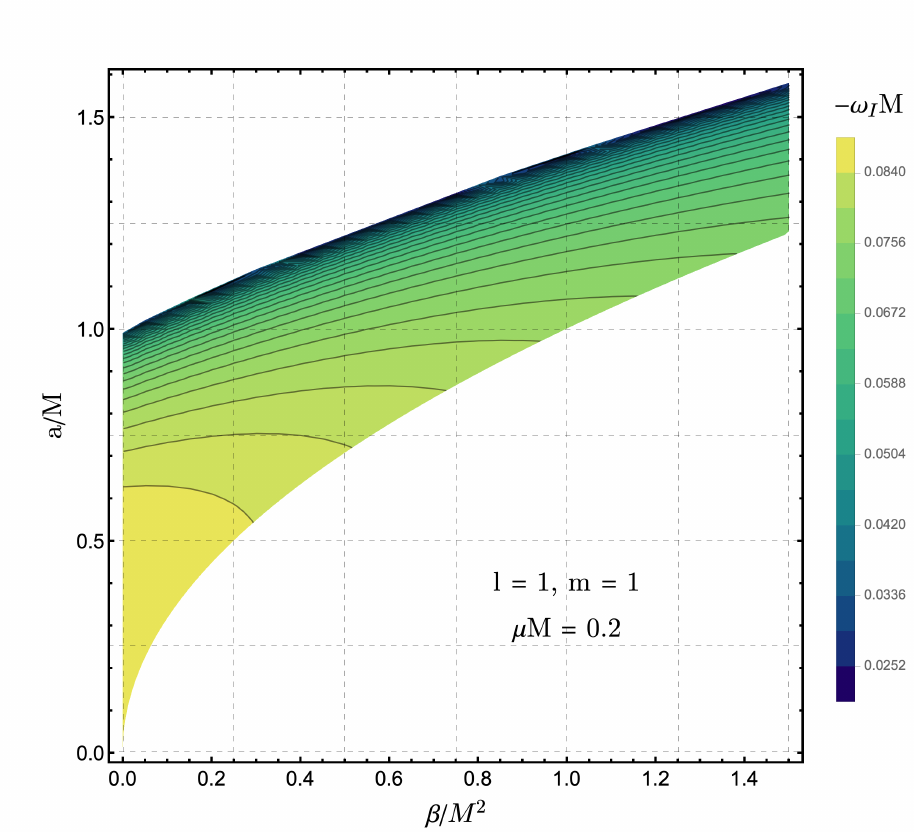}
	\endminipage
	\caption{The imaginary part of the scalar quasinormal mode spectra corresponding to $l=1$ and $m=-1$ (left column), $m=0$ (middle column), $m=1$ (right column) for the allowed values of $a/M$ and $\beta/M^2$ for $\mu M=0$ (top row) and $\mu M=0.2$ (bottom row).}\label{fig:QNM_Imaginary_l1}
\end{figure*}

The fundamental quasinormal modes of the rotating braneworld black hole has been show in \figref{fig:qnm_parametric_plot} for various (normalized) values of the tidal charge $\beta$, BH spin $a$, and scalar field mass $\mu$, each subfigure corresponding to different values of $l$ and $m$. Each curve in the complex plane is labeled by $\beta$, and the color of each point corresponds to the value of $a$, whereas the value of $\mu$ is indicated by the shape of the marker. In \figref{fig:qnm_parametric_plot}, we have restricted ourselves to $0\leq\beta\leq1$, with the value of $a$ being constrained by \eqref{boundsbeta}. However, theoretically there is no upper bound on $\beta$. Hence, we have also explored the $\beta>1$ regime as well, and to get an idea of how the QNMs behave for different values of $\beta, a, \mu$, we have presented contour plots showing how the real and imaginary part of the QNMs vary with respect to $a$ and $\beta$ for massless and massive perturbations in \figref{fig:QNM_RealImg_l0}, and in \figref{fig:QNM_Real_l1} and
\figref{fig:QNM_Imaginary_l1} for $l=0,1$ respectively.

For QNMs, recall that $\mathrm{Re}{(\omega)}$ represents the frequency of oscillation and $\abs{\mathrm{Im}{(\omega)}}$ stands for the rate of decay of the perturbations. For the $l = m= 0$ modes, we observe from the top left panel of \figref{fig:qnm_parametric_plot} that if we fix $\beta$, then the frequency of oscillation increases with $a$ until it reaches its maximum value. After reaching the maximum, the frequency of oscillation decreases with further increase in $a$, while the decay rate becomes nearly constant as the black hole approaches extremality. 

\begin{figure}[!htb]
	\centering
	\includegraphics[width=0.43\textwidth]{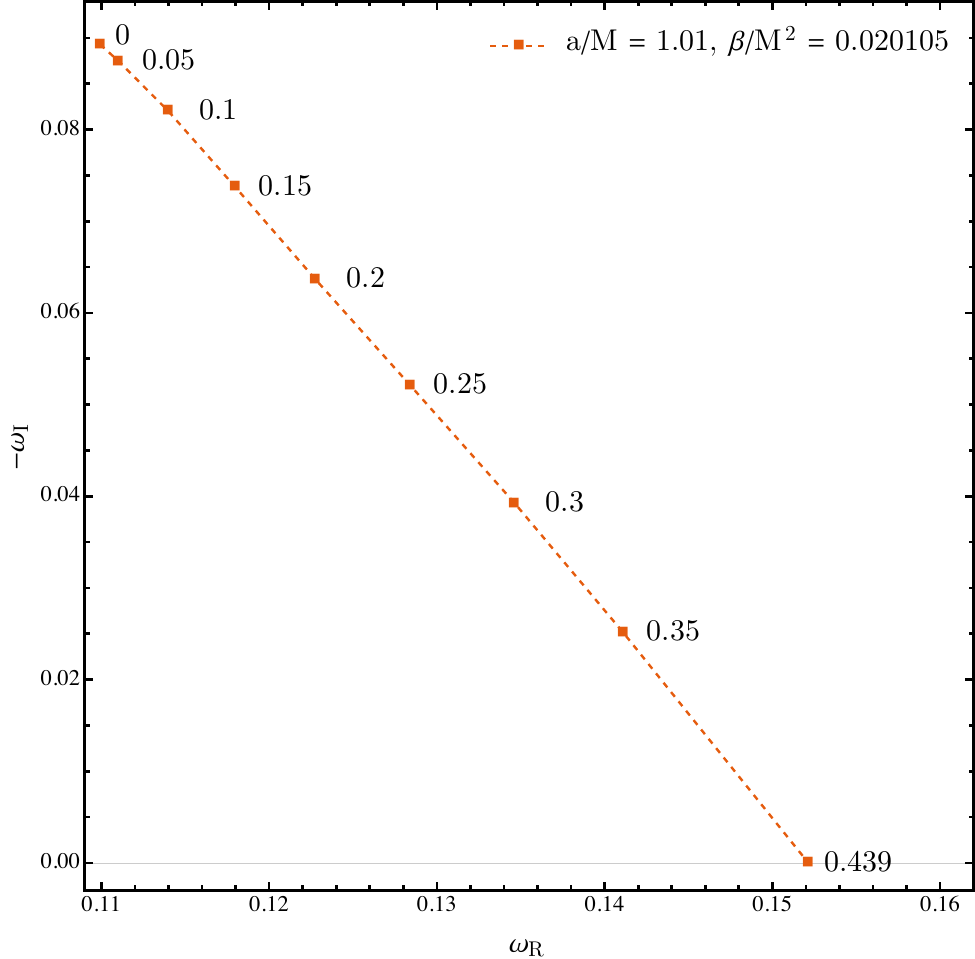}
	\caption{{The formation of quasiresonance state in a near-extremal rotating braneworld black hole spacetime for $l=m=0$ and $a/M > 1$ with $\beta>0$, such that $r_{-}=0.995538 r_{+}$. The labels alongside the points indicate the value of the scalar field mass $\mu M$.}}
	\label{fig:quasiresonance}
\end{figure}

From the top left panel of \figref{fig:qnm_parametric_plot}, we can also see that when the mass $\mu$ of scalar field is turned on, the modes move closer to the real axis. However, in the presence of $\mu$, the change in the decay rate $\mathrm{Im}{(\omega)}$ is much more pronounced than that in the frequency of oscillation $\mathrm{Re}{(\omega)}$. {In fact, if we keep $\beta$ and $a$ fixed, and gradually increase $\mu$, then the imaginary part of the QNM will increasingly tend to a value very close to zero. Since these modes will have a finite $\mathrm{Re}{(\omega)}$ and an extremely small but negative $\mathrm{Im}{(\omega)}$, these modes will be arbitrarily long-lived. Such modes are called \emph{quasiresonance modes} \cite{Ohashi:2004wr,Konoplya:2004wg,Konoplya:2005hr,Zhidenko:2006rs,Dolan:2007mj,Kokkotas:2015uma,Hod:2015goa,Hod:2016jqt,Konoplya:2017tvu,Konoplya:2019hml,Churilova:2019qph,Percival:2020skc,Xiong:2021cth,Rahman:2023swz}, and they satisfy the QNM boundary conditions. They exist even for near extremal configurations, and we explicitly demonstrate the same in \figref{fig:quasiresonance}.}

The contour lines in the top panel of \figref{fig:QNM_RealImg_l0} indicate that the $l=m=0$ modes with the largest oscillation frequency occur for smaller values of the tidal charge (that is, toward the left of the parameter space under consideration) and from the bottom panel of the same figure, we can see that these modes are also associated with a higher decay rate indicating that they are the least long-lived modes. We also see that the oscillation frequency is maximum when the two horizons of the BH are moderately separated but when the scalar field acquires mass, the maximum begins to shift toward extremal configurations. It is also evident from the bottom panel of \figref{fig:QNM_RealImg_l0} that the decay rate of perturbations in extremal configurations are smaller and hence they are more long-lived than their subextremal counterparts. Turning on $\mu$, we can see (from scales on the color bars) the corresponding decay rates become smaller. Lastly, we observe that the long-lived modes lie in the upper-right region of the parameter space for both the massless and massive cases, that is, in the regime of high values of both $\beta$ and $a$, and the corresponding modes have the lowest frequency of oscillation.


Next we consider $l = 1$ modes, and we notice in \figref{fig:qnm_parametric_plot} that the behavior of the modes varies with azimuthal number $m = -1,0,1$.  It is evident from \figref{fig:qnm_parametric_plot} that all the $l=1$ modes have a higher frequency of oscillation compared to the $l=0$ modes for all the $\beta$ and $a$ values that have been considered. Notice that for the $m=0,1$ modes (bottom left and top right panels respectively), if we keep $\beta$ fixed and increase $a$, then the frequency of oscillation increases whereas for $m=-1$ modes (bottom right panel), it decreases. The decay rates for all $m$ values decrease if $a$ is increased while keeping $\beta$ fixed.

Let us first look closely at the $l=1,m=-1$ modes: in the extreme left panel of \figref{fig:QNM_Real_l1}, we see the modes with the smallest frequency of oscillation occur for larger values of both $\beta$ and $a$ as they tend to cluster on the upper right side of the parameter space, and from \figref{fig:QNM_Imaginary_l1}, it is evident that these modes also have the smallest decay rates; the longest-lived modes therefore occur for smaller values $\beta$ and $a$. The inclusion of $\mu$ does not change the qualitative behavior of the modes even though they tend to reduce the values of both the oscillation frequency and decay rate. In fact, from the bottom right panel of \figref{fig:qnm_parametric_plot}, it is clear that the change in the decay rate is much more drastic in the presence of $\mu$ compared to the change in the oscillation frequency. Moreover, for sufficiently large values of $\beta$ and $\mu$ one could possibly obtain quasiresonance modes near extremality.

Now, based on the middle panels of \figref{fig:QNM_Real_l1} and \figref{fig:QNM_Imaginary_l1}, we observed that the $l=1,m=0$ modes with the smallest frequency of oscillation occurs for large values of $\beta$ but smaller values $a$. But the near-extremal modes (especially those occurring near the top right corner of the parameter space) are the ones with the smallest decay rates, and hence they are longest-lived modes. The aforementioned figures along with the bottom left panel of \figref{fig:qnm_parametric_plot} also show that the presence of the mass $\mu$ increases the oscillation frequency of these modes but at the same time makes the modes long-lived by significantly decreasing the decay rate. In fact, for large enough values of $\beta$ and $\mu$, quasiresonance could be achieved by these modes as well.

Let us finally talk about the $l=m=1$ modes in detail: from the extreme right panel of \figref{fig:QNM_Real_l1}, we see that oscillation frequencies of both massless and massive perturbations are higher for black holes near extremality, and since the contours are nearly parallel to the extremal curve, it indicates that they approach a constant value. The smallest values of both $a$ and $\beta$ corresponds to modes with the smallest oscillation frequencies. When it comes to the decay rates, from the extreme right panel of \figref{fig:QNM_Imaginary_l1}, it is clear that the near-extremal modes have the smallest decay rates and hence they are extremely long-lived. In fact, for massless perturbations, they may be extremely close to zero. This feature can also be extrapolated from the trajectories in the complex plane as shown in the top right panel of \figref{fig:qnm_parametric_plot}, and it points toward the existence of the well-known zero damped modes (ZDMs) \cite{Yang:2012pj,Yang:2013uba,Konoplya:2013rxa}. The ZDMs of near extreme and extreme Kerr and KN family of BHs have been extensively studied in the past, and they are quite distinct from the phenomenon of quasiresonance mentioned earlier. The quasiresonance modes are arbitrarily long-lived and have a very small imaginary part: they are associated only with the presence of massive fields whereas ZDMs can occur only for certain values of $m$ for massless perturbations. Now the mass of the scalar field has an intriguing effect on the behavior of the modes: from \figref{fig:qnm_parametric_plot} we see that for smaller values of $\beta$ and $a$, increasing $\mu$  enhances the oscillation frequency slightly and reduces the decay rate but the effect is completely washed out as one approaches extremality. Notice how the modes corresponding to different values of $\mu$ clump together as one increases $a$ while keeping $\beta$ fixed in the top right panel of \figref{fig:qnm_parametric_plot}. It seems likely that near extremality, the quasiresonance frequencies approach those of the ZDMs. However, the interplay between ZDMs and quasiresonance modes needs to be probed further numerically to arrive at a definite conclusion. We wish to return to such questions in the future.

{At this juncture, it would be pertinent to briefly review some of the previous studies related to the quasinormal mode spectra of braneworld black holes and put our present work in its proper context. Notably, massless scalar and gravitational QNMs were studied in \cite{Shen:2006pa, Toshmatov:2016bsb} for the spherically symmetric case. In \cite{Chowdhury:2018pre}, the authors studied massive scalar and Dirac perturbations in a mutated Reissner-Nordstr\"{o}m black hole which is degenerate to the spherically symmetric braneworld solution found in \cite{Dadhich:2000am}. In the parameter space shown in \figref{fig:parameter_space}, these solutions lie along the $x$ axis and have not been considered in the present study as we have focused on rotating solutions with two horizons.}

{The gravitational QNM spectrum for the rotating braneworld BH has been investigated in \cite{Mishra:2021waw} where the authors focused on the $l=m=2,3$ modes. They had considered rotating BHs with both single and double horizons. It is interesting to ask how the behavior of the gravitational modes compare to the scalar modes studied here. However, such a comparison is restricted by the fact that while analyzing gravitational perturbations, one has to implement an approximation to separate the wave equation. Such an approximation is similar in spirit to the one employed in the Kerr-Newman case, and it restricts one to small values of $\beta$ and $a$ \cite{Berti:2005eb}. For (massive) scalar perturbations, we do not have to employ any approximation to separate the wave equation, and hence we are allowed to explore the entire length and breadth of the parameter space.} 

{In the restricted portion of the parameter space explored in \cite{Mishra:2021waw}, it was reported that for a particular value of $\beta$, if one increases $a$, the frequency of oscillation increases while the rate of decay decreases for the $l=m=2,3$ gravitational mode. They further reported that the change in the decay rate in such a situation was smaller compared to the change in the oscillation frequency. Our study confirms that a similar behavior is exhibited by the $l=m=1$ modes. They have also highlighted that for a fixed value of $a$, with an increase in the value of the tidal charge $\beta$, both the real and imaginary part of the $l=m=2,3$ gravitational QNM decreases. The $l=m=1$ scalar perturbations show a similar behavior for small values of $a$, but the behavior is not monotonically decreasing when one consider larger values of $a$. Moreover, in this case as well, the change in the imaginary part of the QNM frequency is smaller than the corresponding change in the real part. We also note that in \cite{Mishra:2023kng}, the authors tried to constrain the value of the tidal charge using gravitational wave data. But it seems that current observations are unable to strongly discriminate between GR and the braneworld scenario.}

Lastly, we end this section with the null result that we could not to find any mode with $\mathrm{Im}(\omega)>0$ (in the region of the parameter space explored in this work) which points toward the stability of the rotating braneworld black hole as far as QNMs are concerned.

\subsection{Quasibound states and superradiant instability} 

\begin{figure*}[!htb]
	\centering
	\minipage{0.33\textwidth}
	\includegraphics[width=\linewidth]{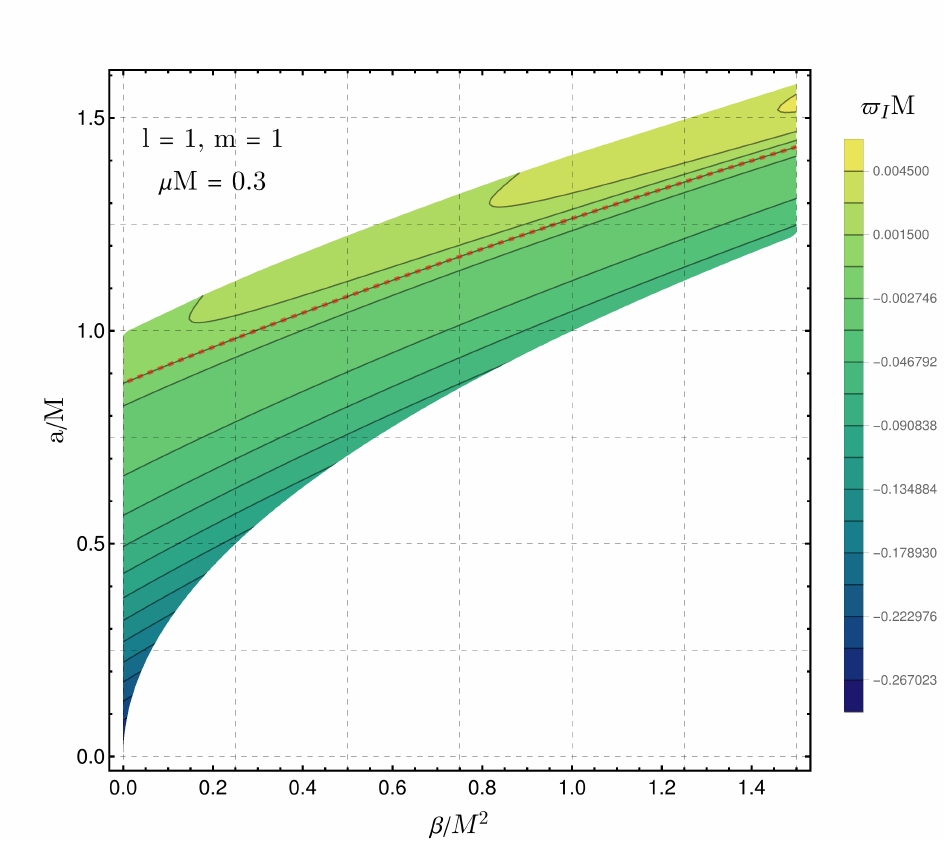}
	\endminipage\hfill
	\minipage{0.33\textwidth}
	\includegraphics[width=\linewidth]{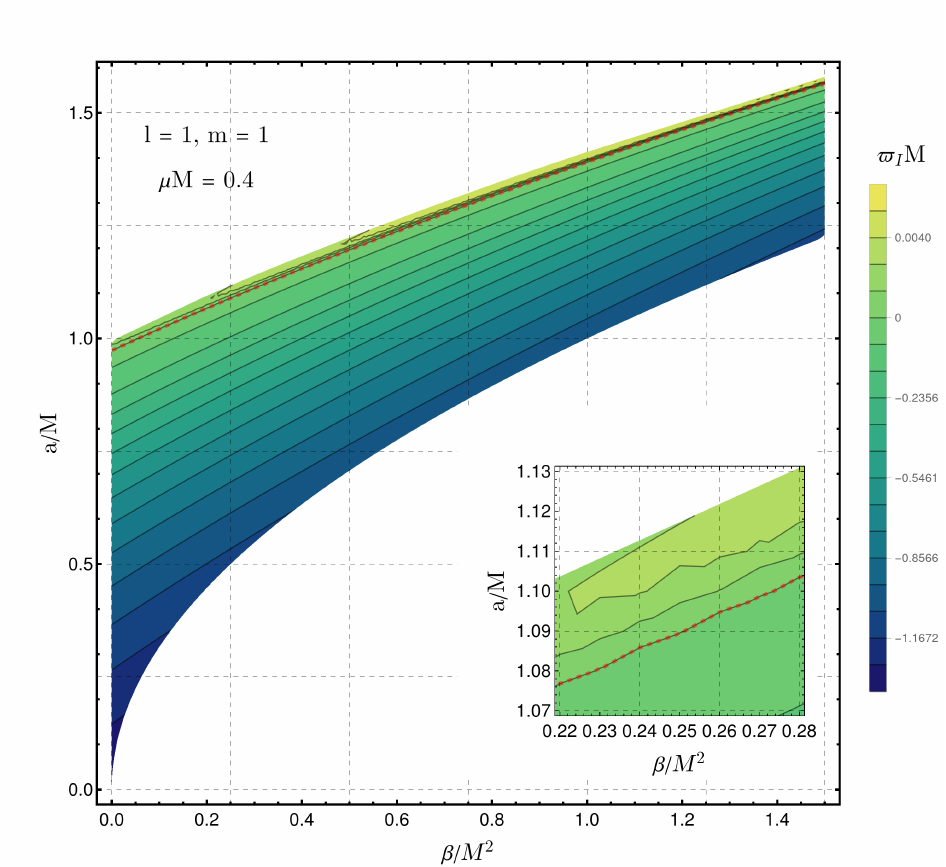}
	\endminipage
	\hfill
	\minipage{0.33\textwidth}
	\includegraphics[width=\linewidth]{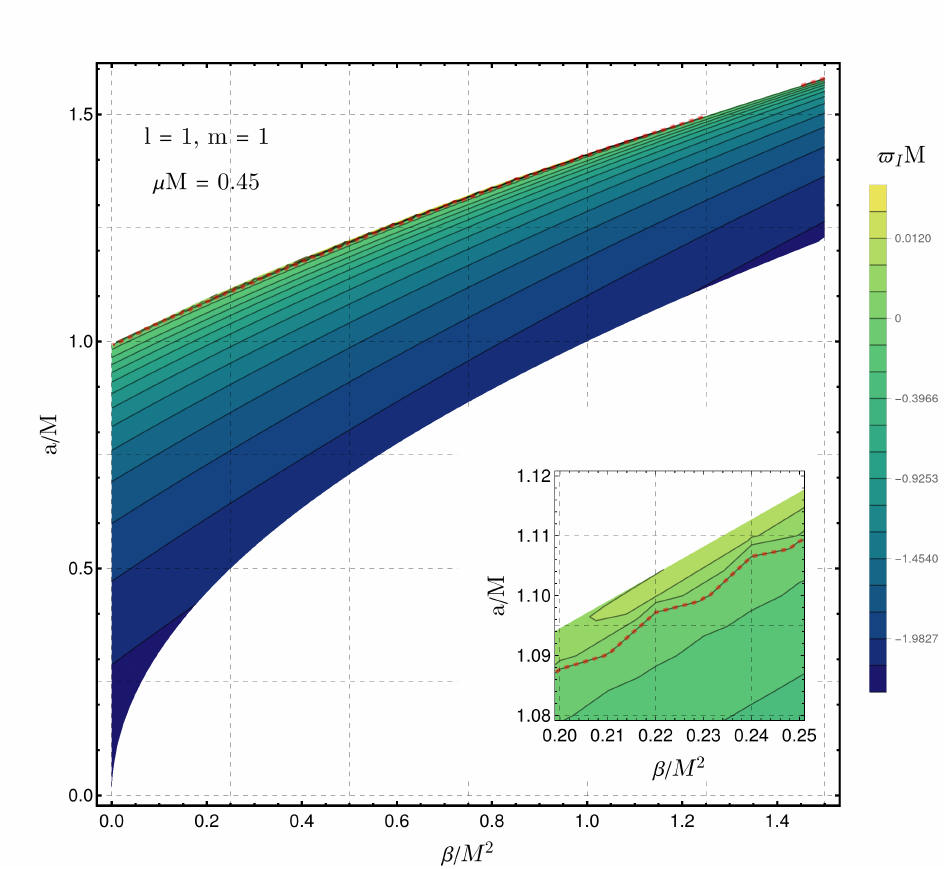}
	\endminipage\hfill
	\minipage{0.33\textwidth}
	\includegraphics[width=\linewidth]{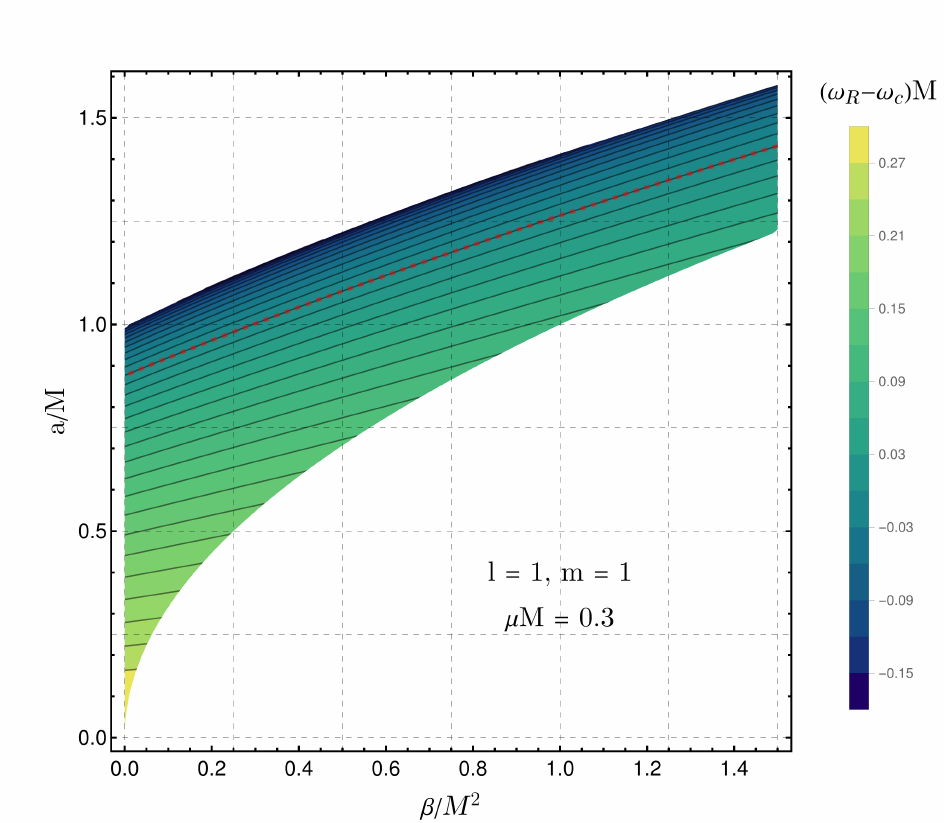}
	\endminipage\hfill
	\minipage{0.33\textwidth}
	\includegraphics[width=\linewidth]{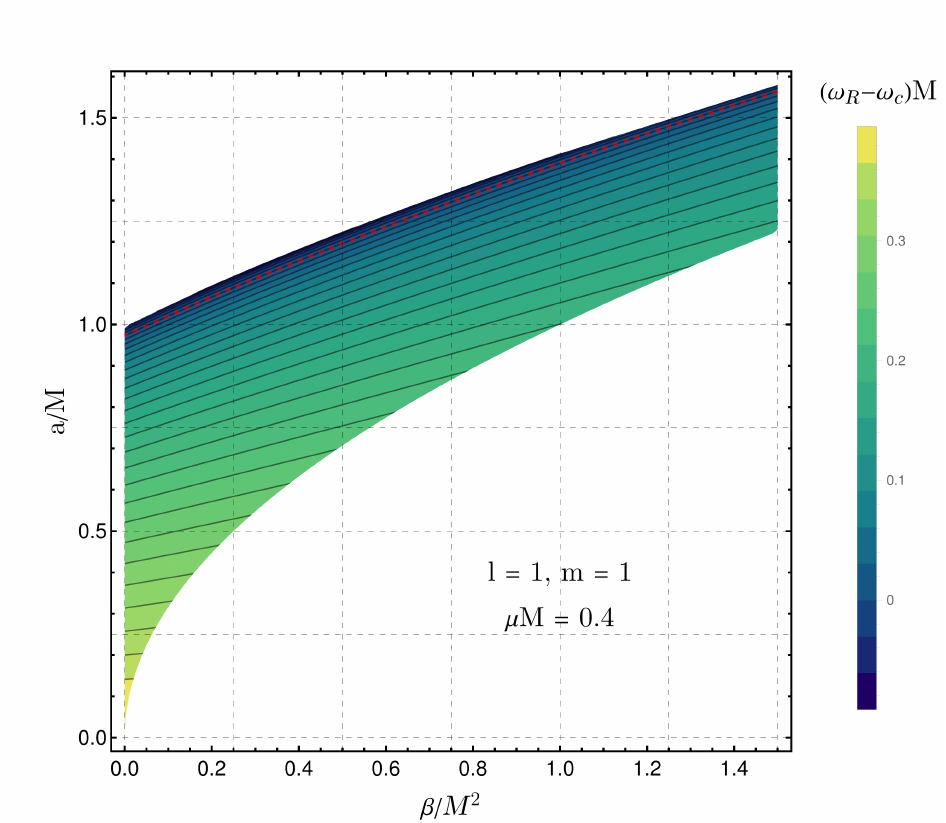}
	\endminipage
	\hfill
	\minipage{0.33\textwidth}
	\includegraphics[width=\linewidth]{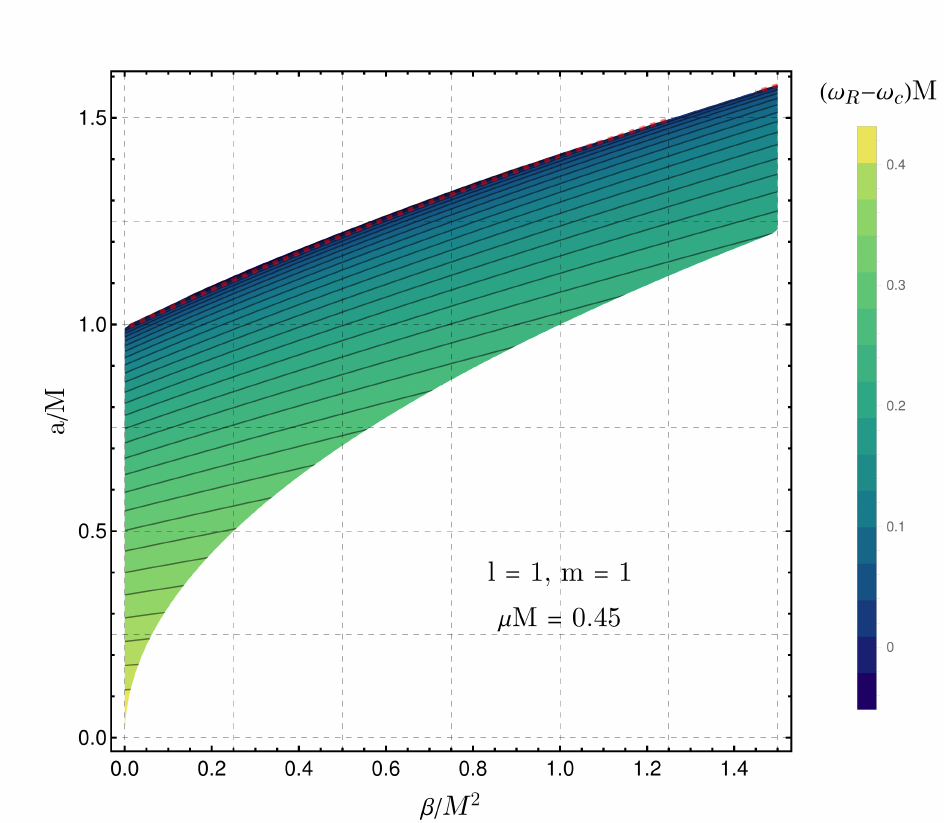}
	\endminipage
	\caption{The imaginary part (top row) of the fundamental quasibound state frequency and the difference between the corresponding real part and $\omega_c=am/(r_+^2+a^2)$ (bottom row) for the $l=m=1$ massive scalar perturbations of mass $\mu M=$ $0.3$ (left column), $0.4$ (middle column), and $0.45$ (right column) of the rotating braneworld black hole. The red dashed curve represents the zero contour line.}\label{fig:CP_superradiant instability}
\end{figure*}

\begin{figure*}[!htb]
	\centering
	\minipage{0.33\textwidth}
	\includegraphics[width=\linewidth]
	{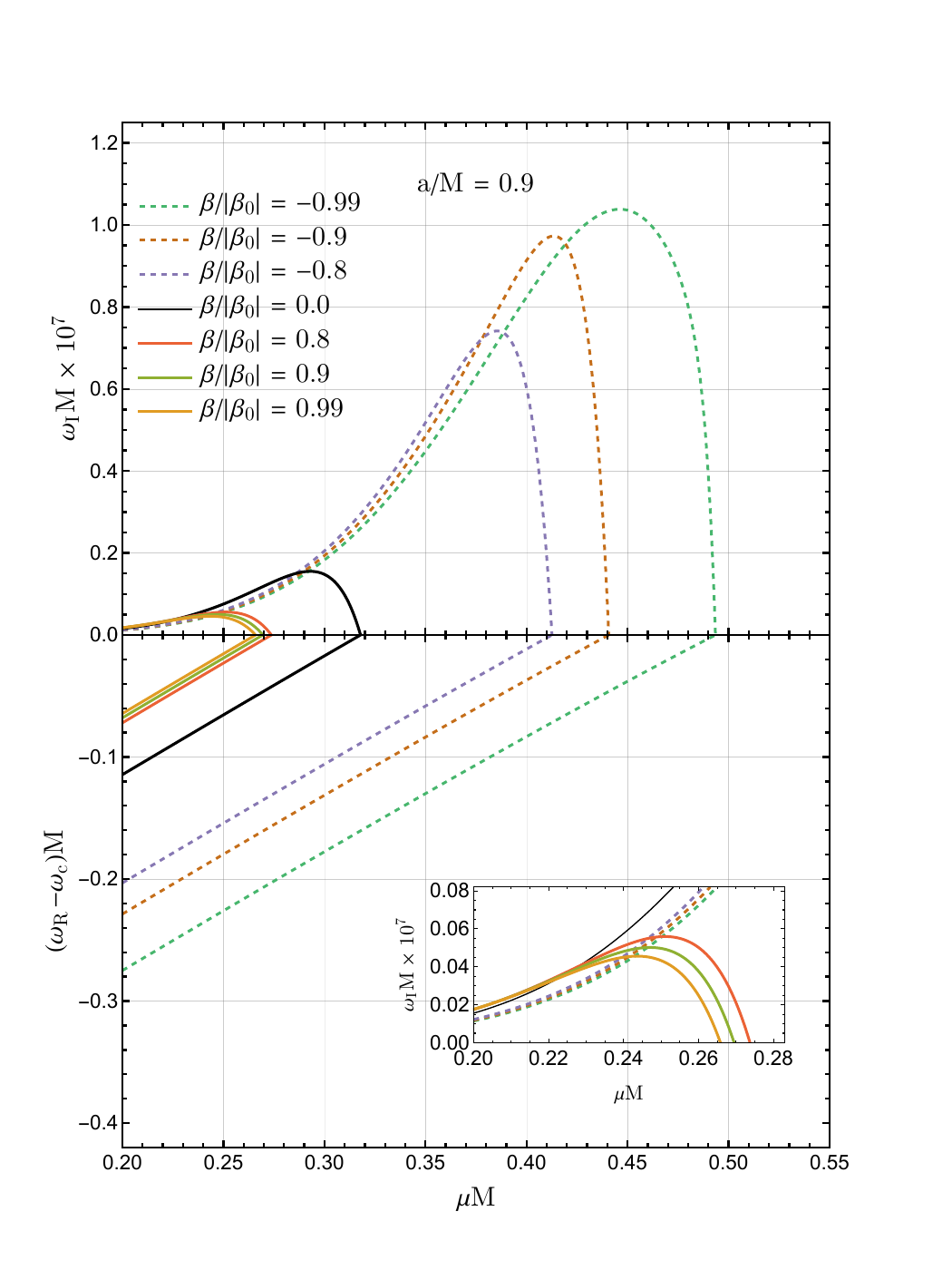}
	\endminipage\hfill
	\minipage{0.33\textwidth}
	\includegraphics[width=\linewidth]{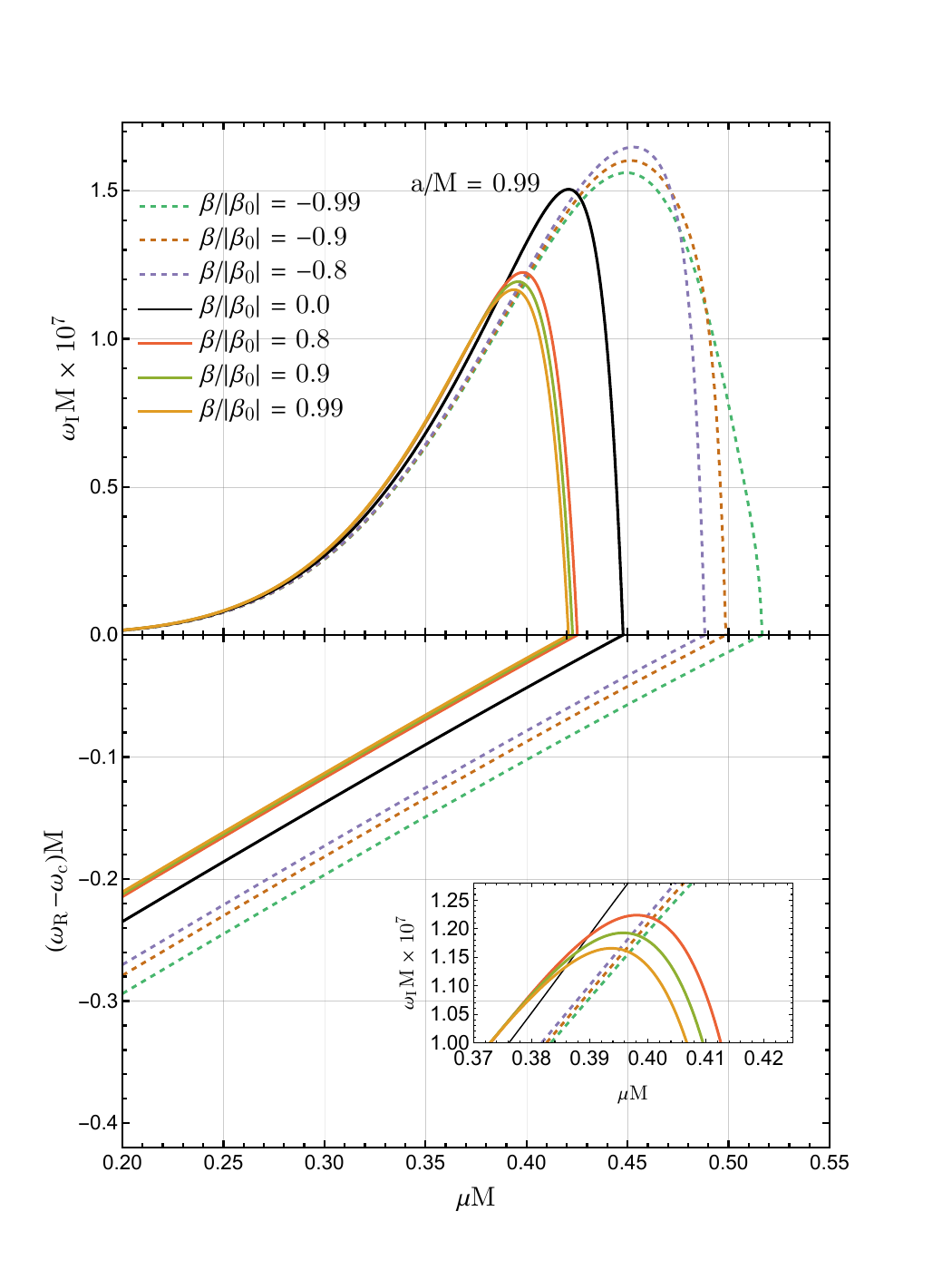}
	\endminipage
	\hfill
	\minipage{0.33\textwidth}
	\includegraphics[width=\linewidth] {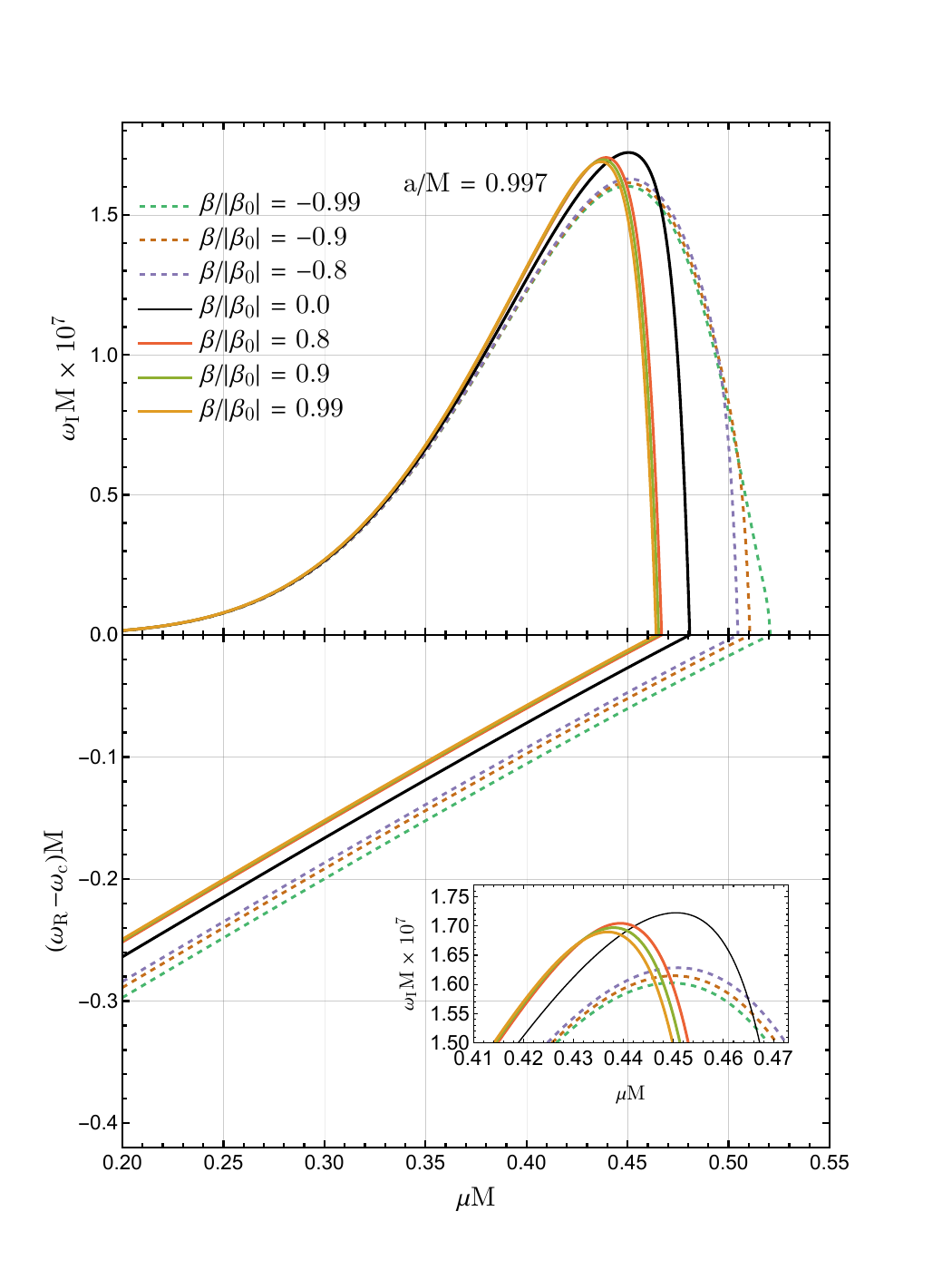}
	\endminipage
	\caption{The superradiant instability associated with the quasibound state spectrum of a massive neutral scalar field ($l=m=1$) for different values of BH spin $a/M<1$ and tidal charge $\beta$.}\label{fig:qbs_a_less_1}
\end{figure*}

\begin{figure}[!htb]
	\centering
	\includegraphics[width=0.45\textwidth]{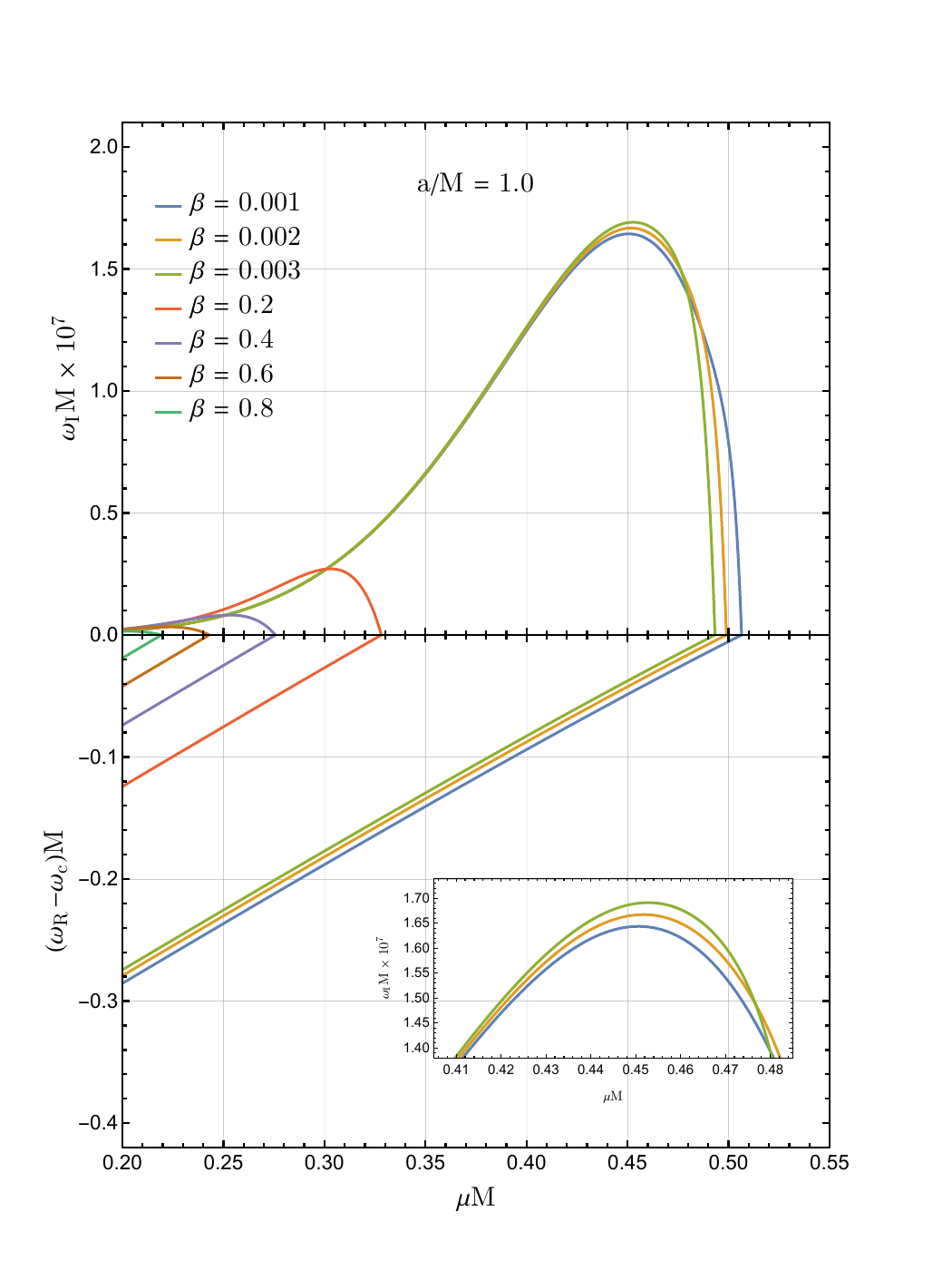}
	\caption{The superradiant instability associated with the quasibound state spectrum of a massive neutral scalar field ($l=m=1$) for $a/M=1$ and different values of the tidal charge $\beta$ (normalized with respect to $M$).}
	\label{fig:qbs_a_1}
\end{figure}

\begin{figure*}[!htb]
	\centering
	\minipage{0.33\textwidth}
	\includegraphics[width=\linewidth]{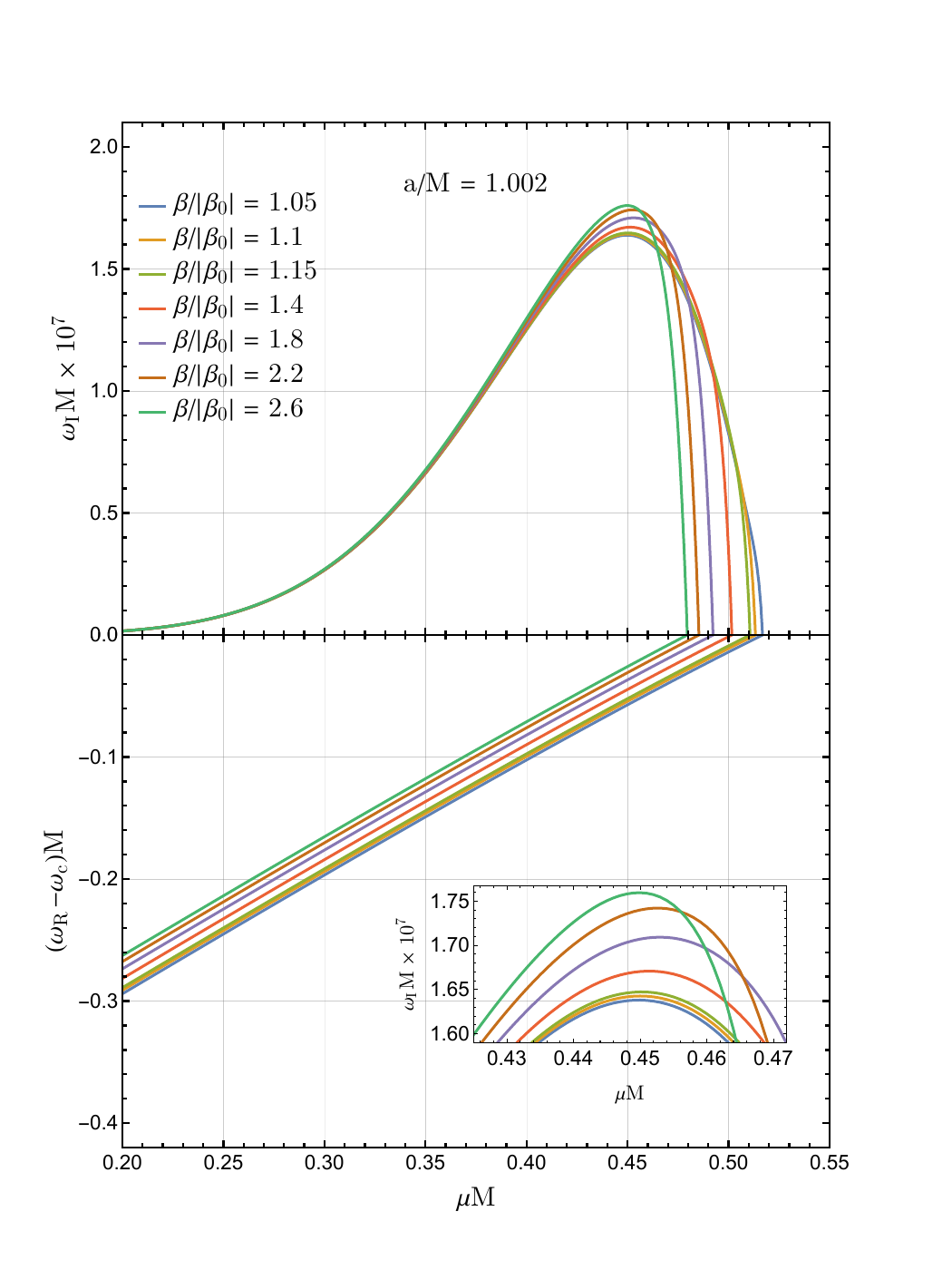}
	\endminipage\hfill
	\minipage{0.33\textwidth}
	\includegraphics[width=\linewidth]{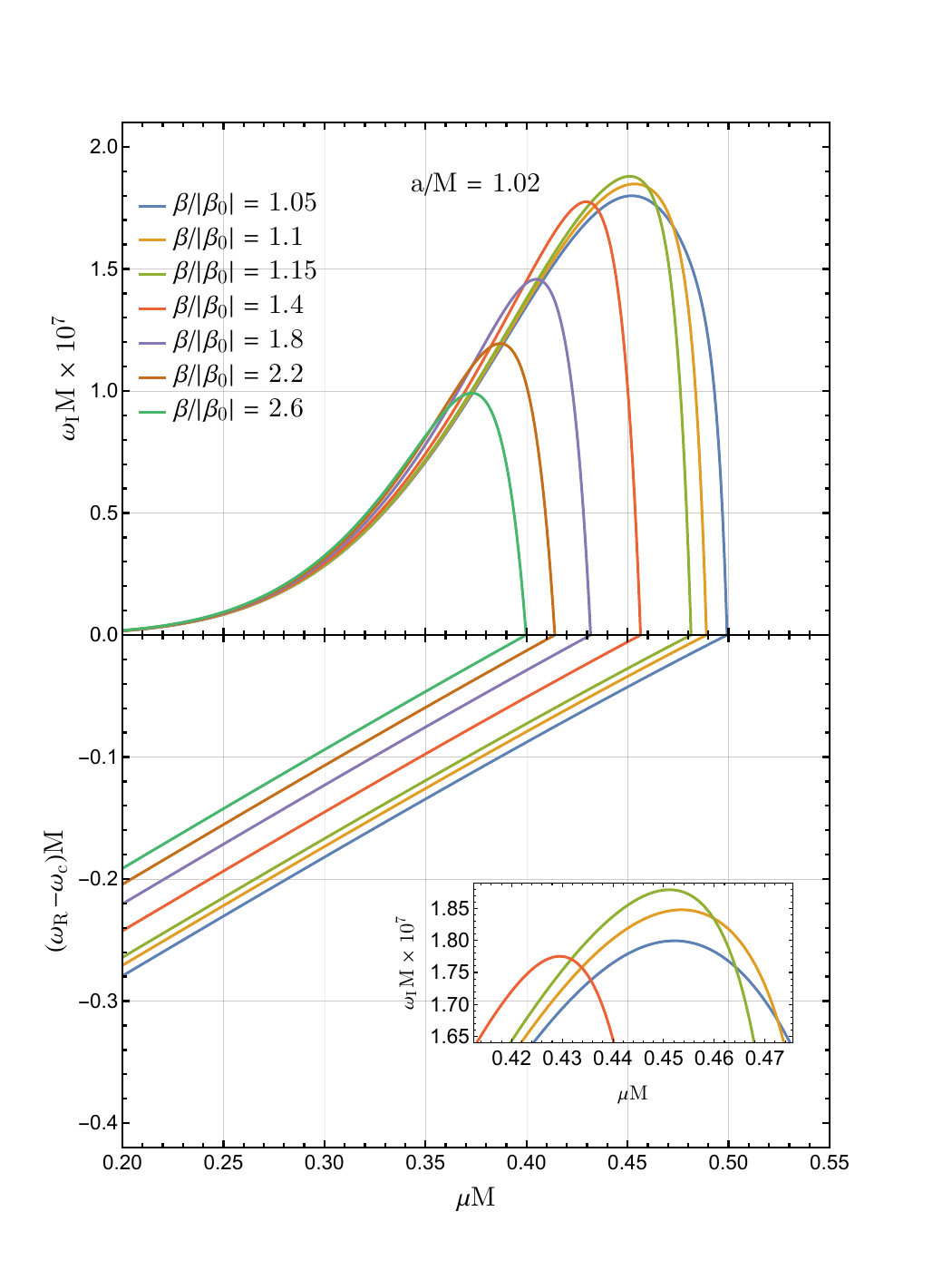}
	\endminipage
	\hfill
	\minipage{0.33\textwidth}
	\includegraphics[width=\linewidth]{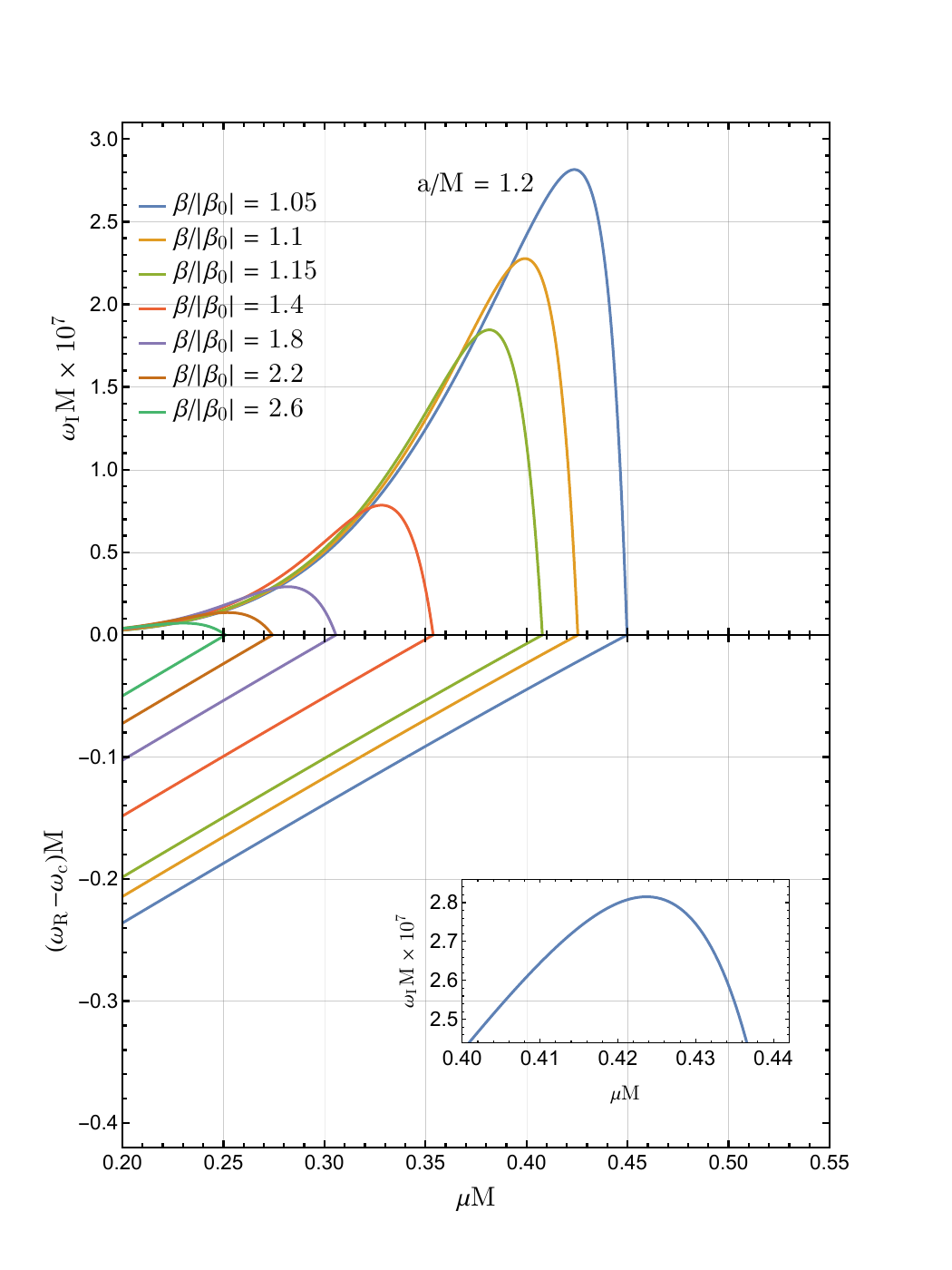}
	\endminipage
	\caption{The superradiant instability associated with the quasibound state spectrum of a massive neutral scalar field ($l=m=1$) for different values of BH spin $a/M>1$ and tidal charge $\beta$.}\label{fig:qbs_a_gtr_1}
\end{figure*}

\begin{figure}[!htb]
	\includegraphics[width=0.45\textwidth]{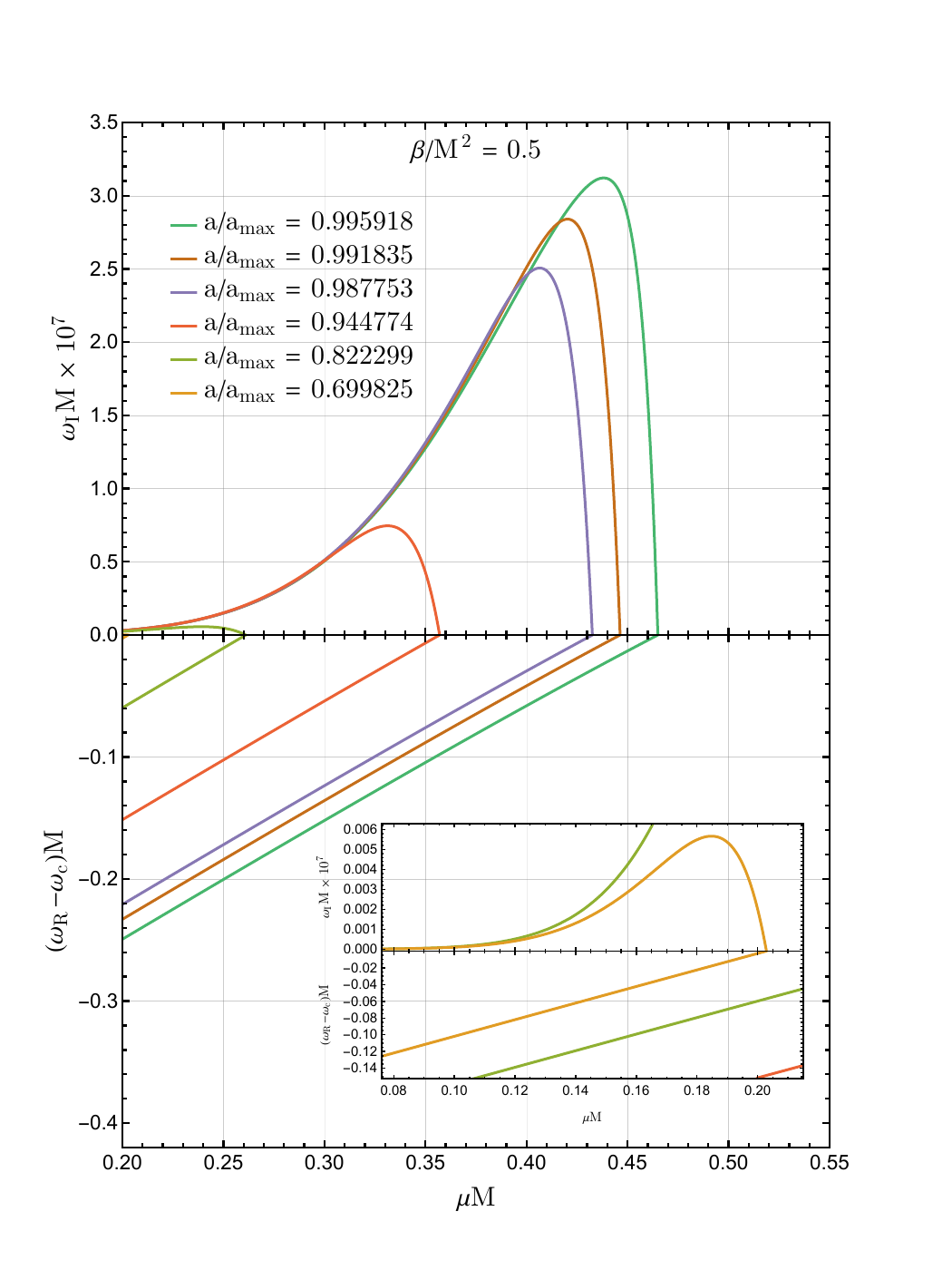}
	\caption{The superradiant instability associated with the quasibound state spectrum of a neutral scalar field ($l=m=1$) for different values of BH spin $a/M>1$ and fixed tidal charge $\beta$.}
	\label{fig:qbs_fixed_beta}
\end{figure}

We now focus on the effect of the tidal charge on the quasibound states (QBSs) and the associated superradiant instability. For massive scalar fields around the Kerr BH, the most unstable modes are those corresponding to $l=m=1$ and $n=0$ {and the maximum superradiant instability occurs for a near extreme Kerr BH when the mass of the scalar field is $\mu M \sim 0.42$ \cite{Dolan:2007mj}. Furthermore, taking a hint from the rich structure of the QBS spectrum of the KN black hole \cite{Huang:2018qdl}, we shall also restrict ourselves to the $l=m=1$ and $n=0$ modes, and first study the quasibound states for three representative values of the scalar field mass, viz., $\mu M = 0.3,0.4,0.45$ over the entire parameter space, the result of which has been shown in \figref{fig:CP_superradiant instability}. We shall then explore the superradiant instability for smaller set of values of $a$ and $\beta$, as shown in \figref{fig:qbs_a_less_1}, \figref{fig:qbs_a_1}, \figref{fig:qbs_a_gtr_1}, and \figref{fig:qbs_fixed_beta}.}

{Before proceeding, we note that the effect of the tidal charge on superradiance was recently studied in \cite{deOliveira:2020lzp} by analyzing the amplification factors of massless scalar waves being scattered by the black hole. However, for massive scalar fields, the analysis has been limited to analytical examinations of extremal configurations \cite{Biswas:2021gvq} where some interesting bounds were obtained on the tidal charge and the BH parameters that ensured that the configuration is superradiantly stable. But a complete numerical analysis of quasibound states and superradiant instabilities of rotating braneworld black hole spanning the entire parameter space has been lacking, and the analysis presented in this section aims to address this gap in the literature.}

{In \figref{fig:CP_superradiant instability}, the top row shows the imaginary part of the QBS frequency and the bottom row shows the difference between the corresponding real part and $\omega_c$ [c.f. \eqref{omega_c}] for three representative values of $\mu M$. The bottom row helps us understand where the superradiance condition given by \eqref{super_cond} is satisfied. Now, the spectrum of quasibound states contain frequencies whose imaginary parts can be positive, negative, or zero. Furthermore, modes that have a positive imaginary part can be very small and are of the order $10^{-7}$ for the Kerr BH \cite{Dolan:2007mj}. So in order to visualize these modes, we use a ``symmetric log" scale, that is, we scaled the imaginary part of the frequency $\omega_I$ as 
$$
\varpi_I=\mathrm{Sgn}(\bar{\omega}_I) \left(\log_{10}(|\bar{\omega}_I M| + 1)\right),
$$
where $\bar{\omega}_I = \mathrm{Im}(\omega) \times 10^{5}$ and $\mathrm{Sgn}$ denotes the signum function. Also, in \figref{fig:CP_superradiant instability}, the red dashed curve corresponds to the zero contour line. From the left column of \figref{fig:CP_superradiant instability}, we see that for $\mu M = 0.3$, the quasibound states with the largest growth rate ($\mathrm{Im(\omega M)>0}$) occurs at the upper right region of the parameter space and hence, is associated with large values of $\beta$ and $a$. These modes also satisfy the superradiance condition. The modes lying on the zero contour indicated by the red dashed line are particularly interesting because they correspond to bound states whose frequencies are purely real with $\mathrm{Re}(\omega) = \omega_c$. The QBSs lying below the zero contour are damped, and the rate of decay increases with the simultaneous decrease of $a$ and $\beta$. Moreover, as the damping increases, the frequency of oscillation increases as well. If we now keep increasing the mass of the scalar field, $\mu M$, the growing modes tend to occur nearer and nearer to the extremal curve \eqref{lowerlimitbeta} as evident from the middle and right columns of \figref{fig:CP_superradiant instability}, and eventually the growing modes will disappear. It is also important to note that the maximum instability for positive values of $\beta$ is of the order of $10^{-7}$, and hence comparable to the Kerr case \cite{Dolan:2007mj}.}

{One can also see that contour lines lying in the region above the zero contour in the top left panel of \figref{fig:CP_superradiant instability} are rather curved compared to the ones lying in the region below. This feature suggests that the BH parameters and $\mu$ together determine the maximum superradiant instability in a rather nontrivial manner. But since \figref{fig:CP_superradiant instability} shows how drastically the region above the zero contour shrinks as one increases $\mu M$ and how the zero contour remains roughly parallel to the extremal curve, focusing on a small set of values of $a$ and $\beta$ near $a/M=1$ is enough to establish the order of the maximum superradiant instability in braneworld black hole. In other words, in the region of the parameter space that we have considered so far, since the contour plots in \figref{fig:CP_superradiant instability} already indicate that the maximum superradiant instability is comparable to that of the Kerr BH and tends to occur near the extremal curve as we progressively increase $\mu M$, we can then focus on a set of values of $a$ and $\beta$ lying around $a/M=1$, without loss of generality, to estimate which values of $\mu M$ can trigger the maximum instability for a set of BH parameters. Such an approach would be perhaps more economical than scanning the entire parameter space for the maximum instability \cite{Siqueira:2022tbc}.}

So, first let us focus on the case when $a/M<1$. The results are summarized in \figref{fig:qbs_a_less_1} where we have considered three values of the rotation parameter, $a=0.9M,0.99M,0.997M$. For each value of $a$ we have considered a set of values of tidal charge $\beta$ normalized by the absolute minimum value of $\beta$, that is, $\beta_0/M^2=\abs{(a/M)^2-1}$ [c.f.:\eqref{lowerlimitbeta}] in order to aid comparison across the $a$ values. Furthermore, the dashed curves in \figref{fig:qbs_a_less_1} correspond to $\beta<0$ which are operationally similar to the KN case discussed in \cite{Huang:2018qdl}, and the solid curves correspond to $\beta>0$. The solid black curve indicates the Kerr case ($\beta=0$). In general, $\mathrm{Im}(\omega M)$ increases with the mass $\mu M$ of the scalar field and becomes positive. It then reaches a maximum value\footnote{We can find the maximum of the curve by constructing an interpolating function using the data generated by the continued fraction method. The function can then be maximized using standard techniques.} for some value of $\mu M$ and on reaching this maximum, it decreases further with increasing $\mu M$ and eventually becomes negative. In the mass range where $\mathrm{Im}(\omega)>0$, $\mathrm{Re}(\omega)<\omega_c$. It is interesting to note from \figref{fig:qbs_a_less_1} that:

\begin{itemize}
	\item for $a=0.9 M$ (left panel of \figref{fig:qbs_a_less_1}), the presence of a relatively large positive tidal charge suppresses the superradiant instability by roughly two orders of magnitude. For $\beta= -0.99 \beta_0$, (i.e., when the BH is near-extreme) we find the highest peak value of $\mathrm{Im}(\omega M)=1.03832 \times 10^{-7}$ at $\mu M=0.446114$ whereas for $\beta=0.99\beta_0$, we observe the smallest peak value $\mathrm{Im}(\omega M)=4.5619 \times 10^{-9}$ at $\mu M=0.243645$. Note that for $\beta= 0$, we find a peak value of $\mathrm{Im}(\omega M)=1.55244 \times 10^{-8}$ at $\mu M=0.293274$.

	\item for $a=0.99 M$ (middle panel of \figref{fig:qbs_a_less_1}), we observe that the peak of the instability does not vary monotonically with $\beta$ \cite{Huang:2018qdl} and the peak occurs for a negative value of the tidal charge (corresponding to a subextreme configuration, in contrast to the previous case). The presence of a positive tidal charge is unable to significantly suppress the superradiant instability. For $\beta= -0.8 \beta_0$, we find the highest peak value of $\mathrm{Im}(\omega M)=1.64681 \times 10^{-7}$ at $\mu M=0.452859$ whereas for $\beta=0.99\beta_0$, we find the smallest peak value $\mathrm{Im}(\omega M)= 1.16555 \times 10^{-7}$ at $\mu M=0.393733$. Note that for $\beta= 0$, we find a peak value of $\mathrm{Im}(\omega M)=1.50435 \times 10^{-7}$ at $\mu M=0.42082$. These results are consistent with \cite{Huang:2018qdl}.

	\item for $a=0.997 M$ (right panel of \figref{fig:qbs_a_less_1}), we obtain the maximum instability of $\mathrm{Im}(\omega M)= 1.72275 \times 10^{-7}$  for $\beta = 0$ at $\mu M=0.450511$ which is again consistent with \cite{Huang:2018qdl}. Interestingly enough, now the smallest positive value of $\beta$ that we had considered has a higher peak than all the negative values of $\beta$. Note that for $\beta= 0.8 \beta_0$, we find the peak value of $\mathrm{Im}(\omega M)=1.70475 \times 10^{-7}$ at $\mu M=0.439384$ whereas for $\beta=-0.99\beta_0$, we find the smallest peak value $\mathrm{Im}(\omega M)= 1.60246 \times 10^{-7}$ at $\mu M=0.449493$.

	\item If one were to solely focus on the positive values of the tidal charge, i.e., $\beta>0$ but $a/M<1$, it is clear from the three insets in \figref{fig:qbs_a_less_1}, that decreasing $\beta$ enhances the superradiant instability, and the corresponding peak values also increases with increase in $a$. But they always remain smaller than the corresponding peak values for $\beta=0$. This appears to be consistent with the findings of \cite{deOliveira:2020lzp} where the author studied the superradiant instability by examining the amplification factors of
	      massless scalar fields scattered by a rotating braneworld BH.

\end{itemize}

Let us now talk about the case when $a/M=1$. The results are visualized in \figref{fig:qbs_a_1}. In this case, the value of $\beta_0=0$ corresponds to the extreme Kerr BH. So we take a few representative values of $\beta/M^2>0$ to study the superradiant instability and do not normalize the value of $\beta$. We do not study the $\beta =0, a/M=1$ case because our numerical method is not equipped to handle extreme BHs. {To study extreme BHs, the modification of Leaver's original method suggested in \cite{Onozawa:1995vu, Richartz:2015saa} might be useful.} We notice that moderate values tidal charge (that is, subextremal configurations) are able to significantly suppress the instability whereas the instability is enhanced in the presence of small positive values tidal charges. But intriguingly, the maximum instability \emph{does not occur} for the smallest value of $\beta$ considered. The maximum instability of $\mathrm{Im}(\omega M)= 1.6912 \times 10^{-7}$ occurs for $\beta = 0.003 M^2$ at $\mu M=0.45278$, and we get the smallest peak value of $\mathrm{Im}(\omega M)= 1.57103 \times 10^{-9}$ for $\beta = 0.8 M^2$ at $\mu M=0.202154$. Note that for $\beta = 0.001 M^2$, the peak value is $\mathrm{Im}(\omega M)= 1.64362 \times 10^{-7}$ at $\mu M=0.45062$.

We now come to the case when $a>1$. In \figref{fig:qbs_a_gtr_1}, we consider three values of $a = 1.002 M,1.02M,1.2M$ and suitably normalized values of the tidal charge $\beta/M^2>0$ as earlier. We summarize our findings below:

\begin{itemize}
	\item for $a=1.002 M$ (left panel of \figref{fig:qbs_a_gtr_1}), the value of the peak of the instability increases with increasing the tidal charge. In particular, the maximum value of the peak $\mathrm{Im}(\omega M)= 1.7595 \times 10^{-7}$ occurs for $\beta=2.6 \beta_0$ at $\mu M=0.449823$. Note that for the instability is the least for $\beta=1.05 \beta_0$ with a peak value of  $\mathrm{Im}(\omega M)= 1.63817 \times 10^{-7}$ at $\mu M=0.449787$.

	\item for $a=1.02 M$ (middle panel of \figref{fig:qbs_a_gtr_1}), we again see that the peak of the instability does not vary monotonically with $\beta$. The maximum peak occurs for and intermediate value of $\beta =1.15\beta_0$ at $\mathrm{Im}(\omega M)= 1.87952 \times 10^{-7}$ for $\mu M=0.451168$.

	\item for $a=1.2 M$ (right panel of \figref{fig:qbs_a_gtr_1}), we notice a behavior opposite to the one encounter for $a=1.002$. We now see that increasing the tidal charge suppresses the superradiant instability by roughly a couple of orders of magnitude. For $\beta= 1.05 \beta_0$, we find the highest peak value of $\mathrm{Im}(\omega M)=2.8151 \times 10^{-7}$ at $\mu M=0.423676$ whereas for $\beta=2.6\beta_0$, we observe the smallest peak value $\mathrm{Im}(\omega M)=7.17095 \times 10^{-9}$ at $\mu M=0.230162$.

	\item if we focus on $\beta = 1.05 \beta_0$ which by design corresponds to a near-extreme BH for all the three values of $a$ that we have considered, we see from the insets in \figref{fig:qbs_a_gtr_1} that the superradiant instability for near extreme BHs with $a/M>1$ intensifies with increase in the value of $a$ (and also the bare value of the tidal charge $\beta$). Looking at the same for  $\beta = 2.6 \beta_0$, the opposite conclusion holds for subextremal braneworld black holes.

\end{itemize}

Lastly, in \figref{fig:qbs_fixed_beta} we keep the tidal charge fixed at $\beta = 0.5 M^2$, and study the superradiant instability by varying $a$ whose values are normalized with respect to $a_\mathrm{max}/M=\sqrt{1+\beta/M^2}$. We note that the superradiant instability intensifies as $a$ increases. In particular, the maximum value of the peak is  $\mathrm{Im}(\omega M)=3.12115 \times 10^{-7}$ at $\mu M=0.438173$ for a near-extreme BH with $a=1.21974 M^2$, or $a=0.995918 a_\mathrm{max}$.

We summarize our results for the maximum peak of the superradiant instability for various values of $a$ and $\beta$ in Table \ref{tab:peak_values}.

\begin{table}[!htb]
	\caption{Maximum peaks of superradiant instability.}
	\begin{ruledtabular}
		\begin{tabular}{cccc}
			$a/M$   & $\beta/M^2$ & $\mu M$  & $\mathrm{Im}(\omega M)$  \\
			\hline                                                      \\
			0.9     & -0.1881000  & 0.446114 & $1.03832 \times 10^{-7}$ \\
			0.99    & -0.0159200  & 0.452859 & $1.64681 \times 10^{-7}$ \\
			0.997   & 0           & 0.450511 & $1.72275 \times 10^{-7}$ \\
			1.0     & 0.003       & 0.45278  & $1.6912 \times 10^{-7}$  \\
			1.002   & 0.0104104   & 0.449823 & $1.7595 \times 10^{-7}$  \\
			1.02    & 0.0464600   & 0.451168 & $1.87952 \times 10^{-7}$ \\
			1.2     & 0.4620000   & 0.423676 & $2.8151 \times 10^{-7}$  \\
			1.21974 & 0.5         & 0.438173 & $3.12115 \times 10^{-7}$ \\ \\
		\end{tabular}
	\end{ruledtabular}
	\label{tab:peak_values}
\end{table}

We end this section with a few comments: First, we note that all of our figures clearly show that instability $\mathrm{Im}{(\omega)}>0$ always occurs in the superradiant regime, $0<\mathrm{Re}(\omega)<\omega_c$. Second, it is in general very difficult to accurately determine for what values of $a,\beta,\mu$, the superradiant instability will be maximum. However, based on our numerical study we may conclude that i) for  $\beta>0$ and $a/M<1$, a higher value of the tidal charge would dampen the instability, ii) for $\beta>0$ and $a/M \geq 1$, the maximum superradiant instability does not vary monotonically with $\beta$ {but the order of the maximum value is the same as that of the Kerr black hole}. The behavior is highly nontrivial, especially when one compares it to results reported previously for massless scalar fields \cite{deOliveira:2020lzp}.

\section{Final Remarks}
\label{remarks}
The RS II rotating braneworld black hole solution provides us with a springboard to test the presence of an extra noncompact spatial dimension on gravitational interactions in the strong field regime. In this study, we have focused on the behavior of a massive scalar field with $\mu M<1$ propagating in the said black hole spacetime. Since the braneworld BH can be superspinning, we have focused on the region of the parameter space where $a/M>1$ and $\beta>0$. First we have made an in-depth study of the quasinormal mode spectra of massive scalar perturbations and have noted the intricate behavior shown by the modes corresponding to different values of the azimuthal number $m$ for $l=1$. {The behavior is qualitatively similar to that of the Kerr (Newman) black hole \cite{Berti:2005eb,Konoplya:2006br,Konoplya:2013rxa}}. We have further explored the formation of quasiresonance modes and discussed the existence of zero damped modes as well.

Next, we have studied the $l=m=1$ quasibound states and superradiant instability associated with such modes, that is, the formation of the so-called gravitational atom. Our analysis reveals a highly nontrivial dependence of the peak of the superradiant instability on the tidal charge and the angular momentum of the black hole. For $a/M<1$, the presence of the tidal charge always dampens the superradiant instability when one compares it to that of the Kerr BH. However, the dynamics is much richer when one looks at superspinning ($a/M>1$) configurations. Notably, for  such near extremal BHs, the superradiant instability intensifies with the tidal charge, although the maximum superradiant instability is comparable to that of the Kerr black hole. These findings could have implications for ongoing efforts to detect boson clouds around black holes in order to constrain the mass of ultra light particles \cite{Arvanitaki:2010sy,Baumann:2018vus,GRAVITY:2023cjt}.

The present work offers numerous extensions. We are attempting to investigate the phenomena of eigenvalue repulsion \cite{Dias:2021yju,Dias:2022oqm} in the quasinormal mode spectra of the rotating braneworld black hole. Moreover, it has been pointed out that the QNM spectrum of black holes may be unstable against small perturbations to the scattering potential \cite{Jaramillo:2020tuu,Sarkar:2023rhp}. It would be worthwhile to study the effect of the tidal charge on the instability of the QNM spectrum. One can also construct braneworld BH solutions which are not asymptotically flat \cite{Neves:2012it} and one may use these solutions to understand how the presence of both the cosmological constant and the extra dimension change the behavior of the QNM spectrum. It would also be interesting to see how the presence of the tidal charge affects the superradiant instability in the regime $\mu \sim \omega$ when the scalar field is charged following \cite{Furuhashi:2004jk,Huang:2018qdl}. Lastly, both scalar and vector boson clouds around Kerr BHs have attracted a lot of attention in recent years, especially in the context of gravitational wave astronomy with binary black holes and people have already explored various aspects of such gravitational atoms \cite{Baumann:2018vus,Baumann:2019ztm,Baumann:2019eav,Baumann:2021fkf,Baumann:2022pkl,Tomaselli:2023ysb,Takahashi:2021yhy,Takahashi:2023flk,Zhang:2018kib,Zhang:2019eid,Tong:2022bbl,Fan:2023jjj}. It would be interesting to extend these studies to the braneworld scenario. Note that in \cite{Cayuso:2019ieu}, the superradiant instability and the formation of the vector gravitational atom was studied for the Kerr Newman black hole. Such a pioneering investigation was possible only after solving the difficult problem of separating the Proca equation by exploiting certain hidden symmetries of the spacetime. Our preliminary investigations indicate that many of these problems can be extended in to the braneworld scenario and the presence of the tidal charge will leave a clear mark of the extra dimension on these systems. We wish to study some of these aspects in the future.


\section*{Acknowledgments} \label{acknowledgment}
The authors would like to thank  Mostafizur Rahman, Srijit Bhattacharjee and Sumanta Chakraborty for useful discussions. S.S. acknowledges funding from SERB, DST, Government of India through the Core Research Grant No. CRG/2020/004347. A.A.S. acknowledges the funding from SERB, Govt. of India under the Core Research Grant No. CRG/2020/004347. S.S.B. acknowledges the funding from the University Grants Commission, Govt. of India under JRF scheme.





%


\end{document}